\documentclass[]{aastex701}

\newcommand{\um}{$\mu$m}
\newcommand{\HII}{\ion{H}{2}~}
\newcommand{\msol}{M\textsubscript{$\sun$}~}
\newcommand{\lsol}{L\textsubscript{$\sun$}~}

\accepted{\date 04 December, 2025}

\submitjournal{ApJ}

\shorttitle{MIRION Catalog of Yellowballs}
\shortauthors{Devine et al.}

\begin{document}

\title{The Milky Way Project: Bridging Intermediate- and High-Mass Star Formation with the \\ MIRION Catalog of Yellowballs}

\author[0000-0002-3723-6362]{Kathryn Devine}
\affiliation{The College of Idaho, 2112 Cleveland Blvd., Caldwell, ID 83605, USA}
\email[show]{kdevine@collegeofidaho.edu}
\correspondingauthor{K. Devine}

\author[0000-0002-9896-331X]{Grace Wolf-Chase}
\affiliation{Planetary Science Institute, 1700 East Fort Lowell, Suite 106, Tucson, AZ 85719 USA}
\email[show]{gwchase@psi.edu}

\author[0000-0003-1539-3321]{C. R. Kerton}
\affiliation{Iowa State University, Department of Physics and Astronomy, 2323 Osborn Dr. Ames, IA 50011, USA}
\email[show]{kerton@iastate.edu}

\author[0000-0002-0828-7726]{Nicholas Larose}
\affiliation{Iowa State University, Department of Physics and Astronomy, 2323 Osborn Dr. Ames, IA 50011, USA}
\email[show]{nrlarose@iastate.edu}

\author{Maya Coleman}
\affiliation{The College of Idaho, 2112 Cleveland Blvd., Caldwell, ID 83605, USA}
\email{fake@gmail.com}

\author{Makenzie Stapley}
\affiliation{The College of Idaho, 2112 Cleveland Blvd., Caldwell, ID 83605, USA}
\email{fake@gmail.com}

\author{Ethan Bassingthwaite}
\affiliation{The College of Idaho, 2112 Cleveland Blvd., Caldwell, ID 83605, USA}
\email{fake@gmail.com}

\author{Bezawit Mekasha Kassaye}
\affiliation{The College of Idaho, 2112 Cleveland Blvd., Caldwell, ID 83605, USA}
\email{fake@gmail.com}

\author{Hritik Rawat}
\affiliation{The College of Idaho, 2112 Cleveland Blvd., Caldwell, ID 83605, USA}
\email{fake@gmail.com}

%CK - Check how Thanrindu would like his affiliation to read. This is based of off a 2025 astro-related paper.
\author[0000-0002-6244-477X]{Tharindu Jayasinghe}
\affiliation{Independent Researcher, San Jose, CA, USA}
\email{unknown@gmail.com}

%\author[0000-0001-9062-3583]{Matthew Povich}
%\affiliation{California State Polytechnic University, Department of Physics and Astronomy, 3801 W. Temple Ave., Pomona, CA 91768, USA}
%\email{mspovich@cpp.edu}

\begin{abstract}

We describe the construction and use of the Mid-InfraRed Interstellar Objects and Nebulae (MIRION) catalog, which was compiled from 6176 objects identified as “yellowballs” (YBs) by participants in the Milky Way Project. The majority of YBs are compact photodissociation regions generated by intermediate- and high-mass young stellar objects that are embedded in star-forming clumps ranging in mass from $10 – 10^6$~\msol and luminosity from $10 – 10^{4}$~\lsol. The MIRION catalog increases the number of candidate intermediate-mass star-forming regions (SFRs) by nearly two orders of magnitude, providing an extensive database with which to explore the transition from isolated low-mass to clustered high-mass star formation. The catalog comprises five tables that include mid- and far-infrared photometry; velocities of source-associated molecular clouds; distances to these molecular clouds; physical properties of source-associated star-forming clumps; and source crossmatches with other catalogs. The structure of the catalog enables users to easily sort objects for further study based on distance or environmental properties. Our preliminary analysis extends our earlier findings that indicate a relationship between IR colors and the physical properties and evolutionary stages of SFRs. Photometry will be periodically updated online to incorporate measurements from volunteers participating in a classroom activity known as the People Enabling Research: a Yellowball Survey of the Colors Of Protostellar Environments (PERYSCOPE) Project. These updates will continue to refine the IR flux measurements and reduce photometric errors. A follow-up paper will present a detailed analysis of how IR colors can be used to predict the properties of star-forming environments.

\end{abstract}

%% \keywords{}

\section{Introduction} \label{sec:intro}

This paper presents the Mid-InfraRed Interstellar Objects and Nebulae (MIRION) Catalog, which contains infrared fluxes, distances, and catalog crossmatches of objects identified as ``yellowballs” (YBs) by participants in the Milky Way Project (MWP) \citep{simpson12}. The MWP was a citizen-science project hosted on the Zooniverse.org platform. Most of the 6176 objects in the MIRION catalog of yellowballs are star-forming regions (SFRs), and evidence indicates that many of these are associated with newly-identified Intermediate-Mass SFRs (IMSFRs) that are producing stars of mass less than 8--10 \msol \citep[][hereafter WKD21]{WC2021}. Therefore, these data observationally probe environments that can help address the question: what causes the transition from low-mass (isolated) to high-mass (clustered) star formation?

Just as solid-state physicists can learn about the nature of materials by studying phase transitions \citep{can2020}, astrophysicists can learn about the nature of star formation by identifying and studying objects that trace an analogous ``phase transition” between low- and high-mass star formation. Previous research efforts exploring this transition have been hampered by the low number ($\sim 50$) of identified IMSFRs (e.g., \citealt{arv2010}). The MIRION catalog fills this observational gap by increasing the number of candidate SFRs producing stars with maximum masses of 3--8 \msol by nearly two orders of magnitude.

MWP participants serendipitously discovered a sample of $\sim 900$ YBs as part of a project aimed at identifying the infrared bubbles/rings in the Milky Way described by \citet{simpson12}. YBs are compact objects whose distinct yellow appearance arises from overlapping 8- and 24-\um~emission, which were assigned green and red colors, respectively, in the Spitzer images used by the MWP. \citet[][hereafter KWA15]{kerton15} showed that many YBs are young, compact photo-dissociation regions (PDRs), which are likely precursors to \HII regions. In this case the 8-\um~emission is produced by UV-excited polycyclic aromatic hydrocarbons (PAHs) and the 24-\um~emission by warm dust.  

KWA15 worked with a YB sample identified by chance. A subsequent iteration of the MWP set YBs as a target \citep{jayasinghe19}, during which volunteers identified over 6000 of these objects from Spitzer surveys covering the inner Galactic plane ($|\ell| \la 65\degr$: \citealt{ben2003,car2009,church2009}), the Cygnus-X complex ($\sim 24$ deg$^2$ centered on $\ell\sim79\fdg3$, $b\sim1\degr$: \citealt{CygX2007,beer2010}), and a portion of the outer Galaxy ($\ell\sim 102\degr - 109\degr$, $b\sim0\degr - 3\degr$: \citealt{smog}). WKD21 published the locations and user-measured radii of these YBs, as well as the results of a pilot-region study of $\sim 500$ YBs in a 20~deg$^2$ region of the Galactic Plane between $\ell = 30\degr - 40\degr$. Details of the construction of the YB database from the raw citizen-science measurements are also presented in WKD21. 

WKD21 identified many more YBs with lower luminosity than those studied in KWA15, and they derived physical properties of YB environments using associations with cores/clumps in the Hi-GAL compact source catalog \citep{Elia2017}. The study demonstrated that YBs identify a mix of young IMSFRs and massive SFRs. This makes these sources particularly interesting, as they highlight the locations of compact (sub-parsec) SFRs that have a wide range of luminosity ($10 – 10^{4}$~\lsol) and mass ($10 – 10^6$~\msol). The YB database expands the number of candidate IMSFRs and identifies many environments thought to be precursors to optically revealed Herbig Ae/Be nebulae.

KWA15 and WKD21 found that many YBs have infrared colors that are distinct from both \HII regions and evolved objects such as PNe. Furthermore, the results from WKD21 provided intriguing evidence suggesting that color trends are associated with the physical properties and evolutionary stages of YB environments, but their small sample was insufficient to draw conclusions that would full describe this relationship. 

The new MIRION catalog includes infrared photometry values, photometric errors, and reliability flags at 8, 12, 24, and 70 \um~for the full set of 6176 YBs identified by MWP volunteers. Together with catalog crossmatches, the new distances we present enable the determination of physical properties (e.g., mass and luminosity) for 3945 catalog objects. A subsequent paper will present the results of analyzing color trends by the physical properties and evolutionary stages of their environments. With this paper, we also rename these sources from YBs to ``MIRION catalog sources," to more appropriately reflect the fact that MIRION sources are heterogeneous and do not represent a single class of object.

The remainder of this paper is organized as follows: Section~\ref{sec:catalog} summarizes the structure and content of the entire MIRION catalog. Details of the infrared photometry are described in Section~\ref{sec:photom}. Section~\ref{sec:distance} explains the methods used to determine the distance to the MIRION sources, and Section~\ref{sec:xmatch} describes the procedure used for catalog crossmatching. We provide preliminary analysis of the catalog's contents in Section~\ref{sec:analysis} and present a summary and conclusions in Section~\ref{sec:conclusions}.

\section{Catalog Content and Structure} \label{sec:catalog}

The MIRION catalog's data are organized into five tables presenting photometry measurements (Table~\ref{tab:mirion-phot}), velocities (Table~\ref{tab:mirion-velo}), distances (Table~\ref{tab:mirion-dist}), physical properties for Herschel-matched objects (Table~\ref{tab:mirion-hcsc}), and catalog crossmatches (Table~\ref{tab:mirion-xmat}). The catalog pairs each object with a unique source identification number that is included in every table to facilitate queries involving multiple properties or associations. The descriptive Tables 1--5 presented in this paper provide information about the form and content of each part corresponding table in the MIRION catalog. In the following paragraphs we provide additional information on the content to facilitate use of the catalog.

Columns 2--8 of Table~\ref{tab:mirion-phot} contain the basic positional and size information of the MIRION sources. Details of the construction of the MIRION catalog from MWP volunteer inputs, including information about the user-measured radius and hit rate parameters, is provided in Section~2 of WKD21. While collecting photometric measurements for this paper we observed that MIRION sources 3035--3101 had a small systematic positional offset (0.35 to 0.1 arcmin between $\ell\sim 101.23\degr - 103.65\degr$). We speculate this offset was introduced during the aggregation process of volunteer observations. Coordinates for these sources have been adjusted to match the actual source position seen in the Spitzer images. The photometric data and flags in Table~\ref{tab:mirion-phot} are discussed in Section~\ref{sec:photom}. 

Section~\ref{sec:photom} describes the challenges facing infrared photometry in the Galactic plane, and the need for multiple measurements of each source to reduce error and uncertainty in the photometric flux measurements presented in the MIRION catalog. A participatory-science experience designed for astronomy students (Section~\ref{PERYSCOPE}) allows users to contribute additional photometry measurements to this work. The ongoing nature of this project means that the MIRION catalog's photometric measurements are continually being refined, and periodic updates will be published to the online catalog. The latest version of the entire catalog is available as an online-only resource \footnote{https://github.com/astrodevine/MIRION}. 

Velocity and distance information for MIRION sources is given in column 2 of Table~\ref{tab:mirion-velo} and column 3 of Table~\ref{tab:mirion-dist}, respectively. The remaining columns of Table~\ref{tab:mirion-velo} provide more detailed information about the type of molecular-line spectrum used (column 4), the various source surveys (columns 5 -- 11), and comparative values from other studies (columns 13 and 14). Section~\ref{sec:velocities} describes the full procedure used to obtain these data. Similarly, columns 4--7 of Table~\ref{tab:mirion-dist} provide comparative distance information from other studies, and columns 8--16 provide input and output values associated with the distance finding techniques detailed fully in Section~\ref{sec:distance_results}.

Table~\ref{tab:mirion-hcsc} lists physical properties of Herschel-detected clumps \citep{Elia2017,Elia2021} associated with MIRION sources. A detailed description of each quantity along with our error propagation technique is provided in WKD21. Table~\ref{tab:mirion-xmat} provides information about crossmatches between MIRION sources and other catalogs of objects related to star formation. The catalog crossmatching technique is described in Section~\ref{sec:xmatch} and the properties of MIRION objects with Herschel clump associations are explored in Section~\ref{sec:analysis}.

%CK For readability using \input command to insert tables.

\begin{deluxetable*}{rlll}
\digitalasset
\tablewidth{0pt}
\tablecaption{MIRION Catalog -- Photometry \label{tab:mirion-phot}}
\tablehead{
\colhead{Number} & \colhead{Units} & \colhead{Label} & \colhead{Explanation}
}
\startdata
1  & --- & ID       & Source identification number \\
2  & deg & GLON     & Galactic longitude \\
3  & deg & GLAT     & Galactic latitude \\
4  & deg & MWPR     & MWP radius \\
5  & deg & e\_GLON  & Uncertainty in Galactic longitude \\
6  & deg & e\_GLAT  & Uncertainty in Galactic latitude \\
7  & deg & e\_MWPR  & Uncertainty in MWP radius \\
8  & --- & HRATE    & MWP hit rate \\
9  & Jy  & F8       & Flux density at 8 $\mu$m \\
10 & Jy  & e\_F8    & Uncertainty in flux density at 8 $\mu$m \tablenotemark{a} \\
11 & Jy  & F12      & Flux density at 12 $\mu$m \\
12 & Jy  & e\_F12   & Uncertainty in flux density at 12 $\mu$m \tablenotemark{a} \\
13 & Jy  & F24      & Flux density at 24 $\mu$m \\
14 & Jy  & e\_F24   & Uncertainty in flux density at 24 $\mu$m \tablenotemark{a} \\
15 & Jy  & F70      & Flux density at 70 $\mu$m \\
16 & Jy  & e\_F70   & Uncertainty in flux density at 70 $\mu$m \tablenotemark{a} \\
17 & --- & N8       & Number of photometric measurements at 8 $\mu$m\\
18 & --- & N12      & Number of photometric measurements at 12 $\mu$m\\
19 & --- & N24      & Number of photometric measurements at 24 $\mu$m\\
20 & --- & N70      & Number of photometric measurements at 70 $\mu$m\\
21 & --- & f\_SAT   & Saturation flag\tablenotemark{b} \\
22 & --- & f\_MULTI & Multiple source flag\tablenotemark{c} \\
23 & --- & f\_NOSRC & No clear source flag\tablenotemark{b} \\
24 & --- & f\_PCONF & Poor confidence in photometry\tablenotemark{b} \\
25 & --- & f\_CEXT  & Highly circular extended source\tablenotemark{c} \\
\enddata
\tablenotetext{a}{Fractional error: standard deviation/flux density}
\tablenotetext{b}{Four-digit integer with each place corresponding a particular band.}
\tablenotetext{c}{Single-digit integer. Set to 1 if flag criteria is met in any band.}
\tablecomments{Table~\ref{tab:mirion-phot} is published in its entirety in the electronic edition of the {\it Astrophysical Journal}.  A descriptive table is shown here for guidance regarding its form and content.}
\end{deluxetable*}

\begin{deluxetable*}{rlll}
\digitalasset
\tablewidth{0pt}
\tablecaption{MIRION Catalog -- Velocities \label{tab:mirion-velo}}
\tablehead{
\colhead{Number} & \colhead{Units} & \colhead{Label} & \colhead{Explanation}
}
\startdata
1 & ---              & ID         & Source identification number \\
2 & $\rm km~s^{-1}$  & VLSR       & Adopted $V_{\rm LSR}$ \\
3 & $\rm km~s^{-1}$  & e\_VLSR    & Uncertainty in adopted $V_{\rm LSR}$   \\
4 & ---              & SType      & Type of spectrum\tablenotemark{a}  \\
5 & $\rm km~s^{-1}$  & VLSR\_G    & $V_{\rm LSR}$ from GRS \\
6 & $\rm km~s^{-1}$  & VLSR\_S    & $V_{\rm LSR}$ from SEDIGISM \\
7 & $\rm km~s^{-1}$  & VLSR\_F    & $V_{\rm LSR}$ from FCRAO OGS \\
8 & $\rm km~s^{-1}$  & VLSR\_T    & $V_{\rm LSR}$ from ThrUMMS \\
9 & $\rm km~s^{-1}$  & VLSR\_DHT  & $V_{\rm LSR}$ from DHT surveys \\
10 & $\rm km~s^{-1}$ & VLSR\_DGT  & $V_{\rm LSR}$ from dense gas tracer (DGT) surveys \\
11 & ---             & DGT        & DGT survey used for column 10 \\
12 & $\rm km~s^{-1}$ & VLSR\_C    & $V_{\rm LSR}$ from \citet{Elia2021} \\
13 & $\rm km~s^{-1}$ & VLSR\_M    & $V_{\rm LSR}$ from \citet{Mege2021} \\
14 & $\rm km~s^{-1}$ & e\_VLSR\_M   & Uncertainty in $V_{\rm LSR}$ from \citet{Mege2021} \\
\enddata
\tablenotetext{a}{SP = single peak, MP:$n$ = multiple ($n$) peaks, LSNR = low SNR, and OUT = outside of molecular line survey coverage.}
\tablecomments{Table~\ref{tab:mirion-velo} is published in its entirety in the electronic edition of the {\it Astrophysical Journal}.  A descriptive table is shown here for guidance regarding its form and content.}
\end{deluxetable*}

\begin{deluxetable*}{rlll}
\digitalasset
\tablewidth{0pt}
\tablecaption{MIRION Catalog -- Distances \label{tab:mirion-dist}}
\tablehead{
\colhead{Number} & \colhead{Units} & \colhead{Label} & \colhead{Explanation}
}
\startdata
1  & --- & ID          & Source identification number \\
2  & kpc & DIST        & Adopted distance \\
3  & kpc & e\_DIST     & Uncertainty in adopted distance \\
4  & kpc & DIST\_C     & Distance from \citet{Elia2021} \\
5  & kpc & DIST\_M     & Distance from \citet{Mege2021} \\
6  & kpc & e\_DIST\_M  & Uncertainty in distance from \citet{Mege2021} \\
7  & --- & STAT\_M     & Status flag from \citet{Mege2021} \\
8  & --- & PFAR        & P$_{\rm far}$ parameter for \citet{Reid2019} distance finder (RDF) \\
9  & kpc & DIST\_R1    & Most probable RDF distance \\
10 & kpc & e\_DIST\_R1 & Uncertainty in most probable RDF distance \\
11 & --- & PINT\_R1     & Integrated probability for most probable RDF distance \\
12 & --- & ARM\_R1     & Name of associated spiral arm for most probable RDF distance\tablenotemark{a} \\
13 & kpc & DIST\_R2    & Second-most probable RDF distance \\
14 & kpc & e\_DIST\_R2 & Uncertainty in second-most probable RDF distance \\
15 & --- & PINT\_R2     & Integrated probability for second-most probable RDF distance \\
16 & --- & ARM\_R2     & Name of associated spiral arm for second-most probable RDF distance\tablenotemark{a} \\
\enddata
\tablenotetext{a}{See \citet{Reid2019} for spiral arm model details.}
\tablecomments{Table~\ref{tab:mirion-dist} is published in its entirety in the electronic edition of the {\it Astrophysical Journal}.  A descriptive table is shown here for guidance regarding its form and content.}
\end{deluxetable*}

\begin{deluxetable*}{rlll}
\digitalasset
\tablewidth{0pt}
\tablecaption{MIRION Catalog -- Herschel-Matched Sources \label{tab:mirion-hcsc}}
\tablehead{
\colhead{Number} & \colhead{Units} & \colhead{Label} & \colhead{Explanation}
}
\startdata
1 & ---                       & ID      & Source identification number                     \\
2 & pc                        & DIAM     & Diameter                                         \\
3 & pc                        & e\_DIAM  & Uncertainty in diameter                          \\
4 & $\rm M_\sun$              & MASS     & Mass                                             \\
5 & $\rm M_\sun$              & e\_MASS  & Uncertainty in mass                              \\
6 & $\rm L_\sun$              & BLUM     & Bolometric luminosity                            \\
7 & $\rm L_\sun$              & e\_BLUM  & Uncertainty in bolometric luminosity             \\
8 & $\rm L_\sun$/$\rm M_\sun$ & LMRAT    & Bolometric luminosity/mass ratio                 \\
9 & $\rm L_\sun$/$\rm M_\sun$ & e\_LMRAT & Uncertainty in bolometric luminosity/mass ratio  \\
10 & K                        & TEMP     & Temperature from graybody fit                                     \\
11 & K                        & e\_TEMP  & Uncertainty in temperature                       \\
12 & ---                      & LRAT     & Luminosity ratio                                 \\
13 & ---                      & e\_LRAT  & Uncertainty in luminosity ratio                  \\
14 & K                        & TBOL     & Bolometric temperature                           \\
15 & K                        & e\_TBOL  & Uncertainty in bolometric temperature            \\
16 & g~cm$^{-2}$              & SIGMA    & Surface density                                  \\
17 & g~cm$^{-2}$              & e\_SIGMA & Uncertainty in surface density                   \\
\enddata
\tablecomments{Columns 2, 4, 6, and 8 contain distance-dependent quantities that have been rescaled using the adopted distances in Table~\ref{tab:mirion-dist}. Columns 10, 12, 14, and 16 contain distance-independent quantities.}
\tablecomments{Table~\ref{tab:mirion-hcsc} is published in its entirety in the electronic edition of the {\it Astrophysical Journal}.  A descriptive table is shown here for guidance regarding its form and content.}
\end{deluxetable*}

\begin{deluxetable*}{rlll}
\digitalasset
\tablewidth{0pt}
\tablecaption{MIRION Catalog -- Catalog Cross-matches \label{tab:mirion-xmat}}
\tablehead{
\colhead{Number} & \colhead{Units} & \colhead{Label} & \colhead{Explanation}
}
\startdata
1 & --- & ID      & Source identification number \\
2 & --- & HIGAL   & Hi-GAL 360 source identifier \\
3 & --- & AGAL    & ATLASGAL source identifier   \\
4 & --- & CORN    & CORNISH source identifier    \\
5 & --- & CORN\_T & CORNISH source type          \\
6 & --- & RMS     & RMS source identifier        \\
7 & --- & RMS\_T  & RMS source type              \\
8 & --- & WISE    & WISE source identifier       \\
9 & --- & WISE\_T & WISE source type             \\
\enddata
\tablecomments{Table~\ref{tab:mirion-xmat} is published in its entirety in the electronic edition of the {\it Astrophysical Journal}.  A descriptive table is shown here for guidance regarding its form and content.}
\end{deluxetable*}

\section{Source Photometry and Flagging} \label{sec:photom}

To conduct photometry and determine infrared colors of the MIRION sources, we follow a procedure similar to the one employed by WKD21. WKD21 used Spitzer Space Telescope images for 8 and 24~\um~flux measurements and WISE catalog images for measurement at 12~\um. 70~\um~measurements were obtained from the Hi-GAL compact source catalog by \citet{Elia2017}. We extended the WKD21 pilot study by collecting photometric data for the full MWP catalog of 6176 YBs, and conducted new photometric measurements of the sources at 8, 12, 24, and 70 \um. Image data used for photometry were obtained at 8~\um~from the Spitzer GLIMPSE survey \citep{ben2003, church2009, https://doi.org/10.26131/irsa405}, 12~\um~from the WISE catalog \citep{wright2010, https://doi.org/10.26131/irsa151}, 24~\um~from the Spitzer MIPSGAL survey \citep{car2009, https://doi.org/10.26131/irsa435}, and 70~\um~from the PACS survey \citep{https://doi.org/10.26131/irsa82}. The Cygnus-X legacy survey \citep{CygX2007, https://doi.org/10.26131/irsa402} provided images for 466 catalog sources at all four wavelengths, and the SMOG legacy survey \citep{smog} provided images for 262 sources at all four wavelengths (S. Carey, private communication 2022). 

\begin{figure}
    \centering
    \includegraphics[width=0.6\textwidth]{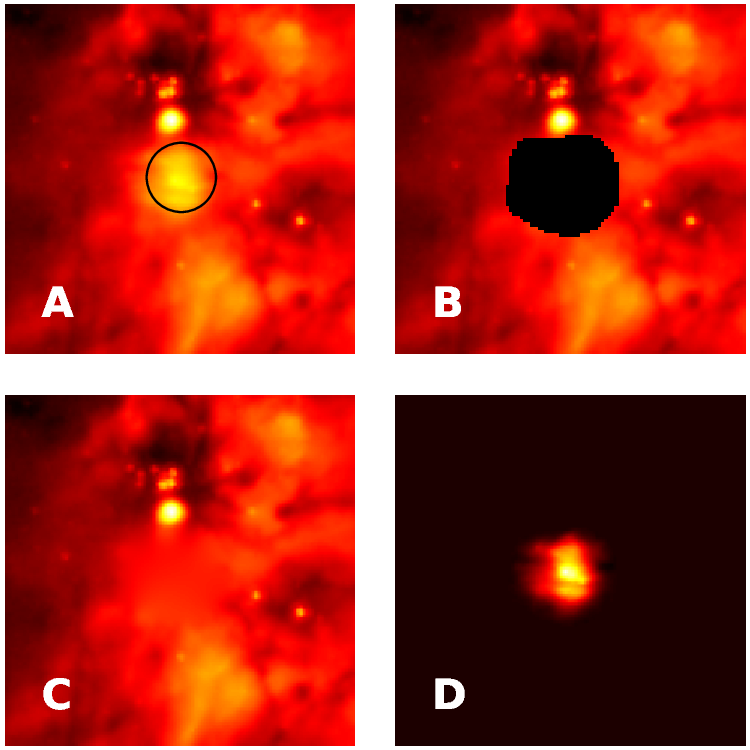}
    \caption{Output images from the Python-based photometry tool used in this work. Upper left panel (A) shows the region containing the source and background emission, where the black circle shows the original MWP average user-selected YB center and radius. Upper right panel (B) shows a user-selected mask covering the source. Bottom left (C) shows the interpolated, background-only image. Bottom t (D) shows the background-removed, source-only image used for photometry.}
    \label{fig:photometry}
\end{figure}

Complex backgrounds in the Galactic plane at MIR wavelengths make automated photometry methods ineffective. WKD21 developed a program to interactively conduct photometric measurements in their pilot region. After this pilot study, the program was modified to additionally include 70~\um~ measurements. Figure \ref{fig:photometry} outlines the photometry process for a representative source. The program loads an image centered on the target object, and the user interactively selects at least three points to generate a polygon that contains the source and distinguishes it from the local background. The program masks the region within these selected points and uses a multiquadratic radial basis function to create a background estimate over the masked area. Finally, the program subtracts this background-only image from the original, resulting in an image of the isolated source, and the flux density is calculated from this ``source-only" image. All of the image files (8, 12, 24, and 70~\um) were re-gridded to match the pixel size of the 8~\um~image to help users be more consistent with the mask region selection process and make the photometry more consistent across wavelengths.

To reduce the uncertainty introduced by variations in the user-based selection of points, %WKD21 collected six measurements of each source at each wavelength. In this study, 
we collected a minimum of five measurements at each wavelength for each source. The number of measurements conducted at each wavelength is listed in Table \ref{tab:mirion-phot}. %The average photometry data for the first four sources in the catalog from these measurements is shown in Table \ref{tab:mirion-phot}. 
Additional measurements will be obtained to further reduce uncertainty in the full catalog through a participatory science outreach effort described in Section \ref{PERYSCOPE}.

%\subsection{Average Mask} \label{AvgMask} %own section or in procedure?

\subsection{Flagging} \label{subsec:flags}

The sources in the MIRION catalog are highly heterogeneous and represent a range of objects. To note complications in the data and interesting morphological features, flags are listed in the catalog in Table \ref{tab:mirion-phot}. These flags were assigned by visual inspection of source-only, background-subtracted images, which were generated using a method similar to that described in the previous section. The flagging code retrieves the coordinates of the masks generated by all of the users who measured the photometry on a given source, and then creates an average mask for the source region by including masked pixels that were selected by at least half of the users who had inspected the source. We then used this average mask to generate source-only images to be examined for flagging. Each source within the catalog was inspected visually and flags were applied. Columns 21--25 of Table~\ref{tab:mirion-phot} show these flags, with 0 indicating no flag and 1 indicating a flag.

\begin{figure}
    \centering
    \includegraphics[width=0.5\textwidth]{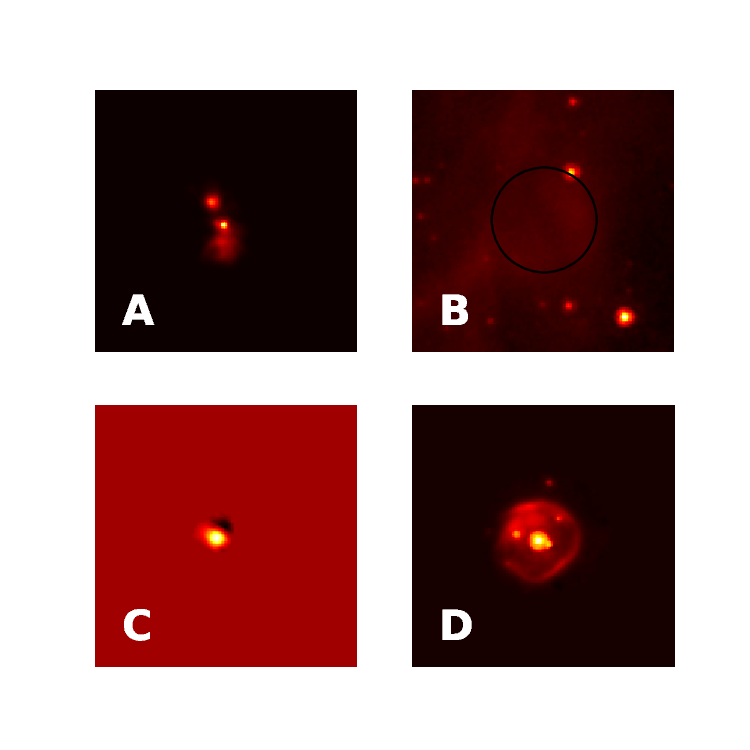}
    \caption{Representative examples of sources to which flags were applied. All images shown are 8 ~\um~background-subtracted, source-only with the exception of (B) which shows the original region since the source-only image which is not meaningful in this case. (A): multiple sources within the masked region. (B): no obvious source at that wavelength. (C): photometry flagged for poor confidence, in which the background subtraction left a negative-valued region near the source, (D): a very round, extended structure.}
    \label{fig:flags}
\end{figure}

Images that are saturated and cannot be used for photometry are denoted by flag ``f\_SAT." Some sources revealed multiple point sources, particularly at 8~\um, denoted by flag ``f\_MULTI." Multiple sources often appear in the 8~\um~image due to the higher resolution at this wavelength, which reveals filaments or other complex features that appear to be simple, circular sources at longer wavelengths. Some sources that appeared ``yellow'' in the MWP images used to generate the original YB database do not have obvious emission at all 8, 12, 24, and 70~\um~wavelengths, denoted with the ``f\_NOSRC'' flag. Sources for which the photometry methods failed to yield reliable results (e.g. sources appearing against a highly complex background) are flagged with a ``f\_PCONF" flag.  Finally, a small number of sources presented interesting, very round structures that are likely planetary nebulae or compact bubbles. These sources are flagged with a ``f\_CEXT" flag. Figure~\ref{fig:flags} shows example sources to which flags were applied. While most of the flags designate potential issues with the photometry, f\_MULTI and f\_CEXT sources are flagged as sources of interest for follow-up studies. For example, some ``f\_MULTI" sources may represent star-forming regions just at the evolutionary stage where their stellar content is being revealed at shorter MIR wavelengths. There are 1938 sources in the MIRION catalog sources are flagged with f\_MULTI, and 18 are flagged with f\_CEXT.

While the f\_MULTI and f\_CEXT flags are not wavelength dependent, the f\_SAT, f\_NOSRC, and f\_PCONF flags may be required at some wavelengths and not others. Thus, these three flags are all represented as four-digit strings for flags at 8, 12, 24, and 70~\um~respectively. For example, a flag f\_NOSRC=0011 indicates no obvious source at 24 and 70~\um, while a flag f\_NOSRC=1000 indicates no obvious source at 8~\um. A f\_PCONF=0100 flag would indicate poor confidence in the 12~\um~photometry for that source, and a f\_SAT=0010 flag would indicate that the 24~\um~image is saturated for that source.

\subsection{Participatory Science through PERYSCOPE} \label{PERYSCOPE}

To collect more photometric measurements and thereby reduce the uncertainty in our photometry measurements, we have developed the People Enabling Research: A Yellowball Survey of the Colors of Protostellar Environments (PERYSCOPE) Project. The PERYSCOPE Project is aimed primarily towards introductory-level astronomy students, but can be adapted to any interested group of volunteers. PERYSCOPE includes an activity guide, instructional videos, and a beginner-friendly version of the photometry code developed using Google Colab that allows participants to learn how astronomers collect and use photometry data in their research. 

PERYSCOPE participants are expected to collect and analyze data on a set of around 40 YBs over one or two class or lab periods. Participants are encouraged to submit their data to our database and are given the option to be acknowledged in publications where their measurements are used. PERYSCOPE user measurements will be used to refine the photometric measurements, and will be periodically incorporated into the online version of the MIRION catalog. User-generated data will be compared to the photometry published with this paper for quality control prior to inclusion in the online catalog. Photometric measurement consistency and overlap of the user-selected source region with the average selected region will be examined as inclusion criteria.

\section{Distances}
\label{sec:distance}
\subsection{Data}
\label{sec:dist_data}
To obtain velocity and distance estimates for MIRION sources, we used the CO molecule and its isotopes (e.g. $^{12}$CO, $^{13}$CO), as these sources are expected to form within molecular clouds. The MIRION sources are categorized into four regions based on their spatial coverage: the first Galactic quadrant (Q1), the fourth Galactic quadrant (Q4), SMOG and Cygnus-X. The primary survey used for Q1 is the Boston University-Five College Radio Astronomy Observatory Galactic Ring Survey (hereafter GRS, \cite{Jackson2006}). The GRS covers a majority of MIRION sources within Q1, with complete coverage from l=18$^\circ$-55.7$^\circ$, and partial coverage down to l=14$^\circ$. It makes use of the $^{13}$CO (J=1$\rightarrow$0) rotational transition, has a spectral resolution of 0.2 km s$^{-1}$, and has a velocity coverage from -5 to 135 km s$^{-1}$ for l$\leq$40$^{\circ}$, -5 to 85 km s$^{-1}$ for l$>$40$^{\circ}$. 

On its own, the GRS does not have complete coverage for Q1; consequently, we supplemented GRS with data from the Structure, Excitation and Dynamics of the Inner Galactic Interstellar Medium survey (SEDIGISM, \citep{Schuller2021}). SEDIGISM observed $^{13}$CO and C$^{18}$O(J=2$\rightarrow$1) from l=-60$^\circ$ to 18$^\circ$, with a spectral resolution of 0.25 km s$^{-1}$. While this survey contains most of the MIRION sources within Q4, we used the ThrUMMS survey \citep{Barnes2015} to extend coverage. ThruMMS observed $^{12}$CO, $^{13}$CO, and C$^{18}$O (J=1$\rightarrow$0) from l=299$^\circ$ to 359$^\circ$, with a spectral resolution of 0.3 km s$^{-1}$. For our purposes, $^{13}$CO was used for consistency with the other surveys. Both the SMOG and Cygnus-X regions lie outside of the high-resolution CO surveys previously mentioned. For a majority of the SMOG region (from $l>103\fdg5$), we used $^{12}$CO $(J=1 \to 0)$ data, with 100\farcs44 spatial resolution and 0.82 km~s$^{-1}$ velocity resolution, from the Canadian Galactic Plane Survey (CGPS) \citep{Taylor2003}. These data are re-processed and re-projected versions of data from the Five College Radio Astronomy Observatory (FCRAO) $^{12}$CO $(J=1 \to 0)$ Outer Galaxy Survey (OGS; \citet{Heyer1998}). Details of the incorporation of the OGS into the CGPS are given in \citet{Brunt2003} and \citet{Taylor2003}.

For the remaining MIRION objects that were not found within any of the aforementioned surveys, we selected low-resolution CO surveys from the 1.2 meter CO Survey Archive. We utilized eight regions from those available (\citet{Dame1987}, \citet{Grabelski1987}, \citet{Bronfman1989}, \citet{Leung1992}, \citet{Bitran1997}, \citet{Dame2001}) to estimate peak CO velocities, using the higher-resolution surveys before resorting to the low-resolution ``Superbeam" survey \citep{Dame1987}. A summary of the CO surveys used, including their observed lines, spatial resolutions, and spectral resolutions, is presented in \autoref{tab:SpectralLineSurveys}. 

In an attempt to disambiguate sources with multiple similarly intense peaks (see \autoref{sec:velocities}), we compared spectra from surveys that trace dense gas or evidence of star formation with CO spectra. MIRION objects are thought to primarily be star-forming regions, and therefore may exhibit signatures of early- or late-stage star formation. Common tracers include maser emission from various molecules, such as OH, CH$_3$OH, and H$_2$O. In order to match as many MIRION sources as possible to potential maser sites, we selected large-scale surveys. These surveys include THOR, The HI/OH/Recombination line survey of the Milky Way \citep{Beuther2019}; SPLASH, the Southern Parkes Large-Area Survey in Hydroxyl \citep{Dawson2022}; and HOPS, the H$_2$O southern Galactic Plane Survey \citep{Walsh2008}. SPLASH provided supplemental data from OH maser emission at 1612, 1665, 1667 and 1720 MHz, while HOPS provided H$_2$O maser emission at 22 GHz and NH$_3$ (1,1) emission, which traces dense gas in star-forming regions.

% NL: 22.235 GHz 
\begin{deluxetable}{lccc}
\tablecaption{Spectral Line Surveys \label{tab:SpectralLineSurveys}}
\tablewidth{0pt}
\tabletypesize{\normalsize}
\tablehead{\colhead{Survey}  & \colhead{Observed Line} & \colhead{Spatial Resolution}       & \colhead{Spectral Resolution} \\
        \colhead{Name} & \colhead{}      & \colhead{(degrees)}           & \colhead{(km s$^{-1}$)}
}
\startdata
FCRAO GRS & $^{13}$CO (J=1$\rightarrow$0) & 0.013 & 0.20 \\
SEDIGISM & $^{13}$CO (J=2$\rightarrow$1) & 0.008 & 0.25 \\
ThrUMMS & $^{13}$CO (J=1$\rightarrow$0) & 0.018 & 0.30 \\
FCRAO SMOG & $^{12}$CO (J=1$\rightarrow$0) & 0.028 & 0.82 \\
DHT01 & $^{12}$CO (J=1$\rightarrow$0) & 0.500 & 0.65 \\
DHT02 & $^{12}$CO (J=1$\rightarrow$0) & 0.125 & 1.30\\
DHT08 & $^{12}$CO (J=1$\rightarrow$0) & 0.125 & 0.65 \\
DHT10 & $^{12}$CO (J=1$\rightarrow$0) & 0.125 & 0.65 \\
DHT17 & $^{12}$CO (J=1$\rightarrow$0) & 0.125 & 0.65 \\
DHT18 & $^{12}$CO (J=1$\rightarrow$0) & 0.250 & 0.65 \\
DHT33 & $^{12}$CO (J=1$\rightarrow$0) & 0.125 & 1.30 \\
DHT36 & $^{12}$CO (J=1$\rightarrow$0) & 0.125 & 1.30 \\
\enddata
\tablecomments{Surveys used in determining velocities of MIRION objects. The low-resolution survey names (DHT\#) are as they appear in the 1.2 meter CO Survey Archive.}% The spatial/spectral resolutions presented for the low-resolution surveys reflect interpolated values when suggested.} 

\end{deluxetable}

\subsection{Velocities}
\label{sec:velocities}
We retrieved velocity estimates from CO survey spectral-cube cutouts (0.$'$64$^2$ in size) at the location of the MIRION sources. The cutout size is based on the average source size being $\sim$ $24''$ across (WKD21). This ensures enough of the source is selected for a sufficient signal-to-noise ratio (SNR), but is not too large as to contain unrelated emission. Comparison of a sample of MIRION source velocities to those adopted by WKD21 using a $24''$ search region size resulted in an average percent difference of 7.7$\%$. Increasing this to $48''$ resulted in a smaller percent difference of 2.1$\%$.%, suggesting the use of a larger cutout region. 
Some of the largest MIRION sources from WKD21 reach projected radii of $\sim 50''$, meaning selecting a cutout region larger than this would result in unrelated emission for a majority of sources. With these considerations in mind, we adopted a $48''$ search region for velocity estimation. It is important to note that while most MIRION sources do not lie within close proximity to one another (with average separation $> 200''$), there are a few instances of multiple sources residing within the same cutout region. In these cases, we assumed that these objects lie within the same parent structure/molecular cloud, and would therefore share similar line-of-sight radial velocities. 

Once we obtained a cutout, we adopted the velocity corresponding to a Gaussian fit about the most intense CO peak (I$\textsubscript{max}$) for each object, as this often traces the most dense gas \citep{Ellsworth-Bowers2015}. Peaks were identified as exceeding the ``noise floor," defined as $<I>$+3$\sigma_{I}$ for 30 channels near the end of the spectrum. To distinguish between secondary peaks and noise associated with a single major peak, peaks were required stand out from nearby minor peaks in excess of 0.2I$\textsubscript{max}$. 

\begin{figure}
    \centering
    \includegraphics[width=0.75\linewidth]{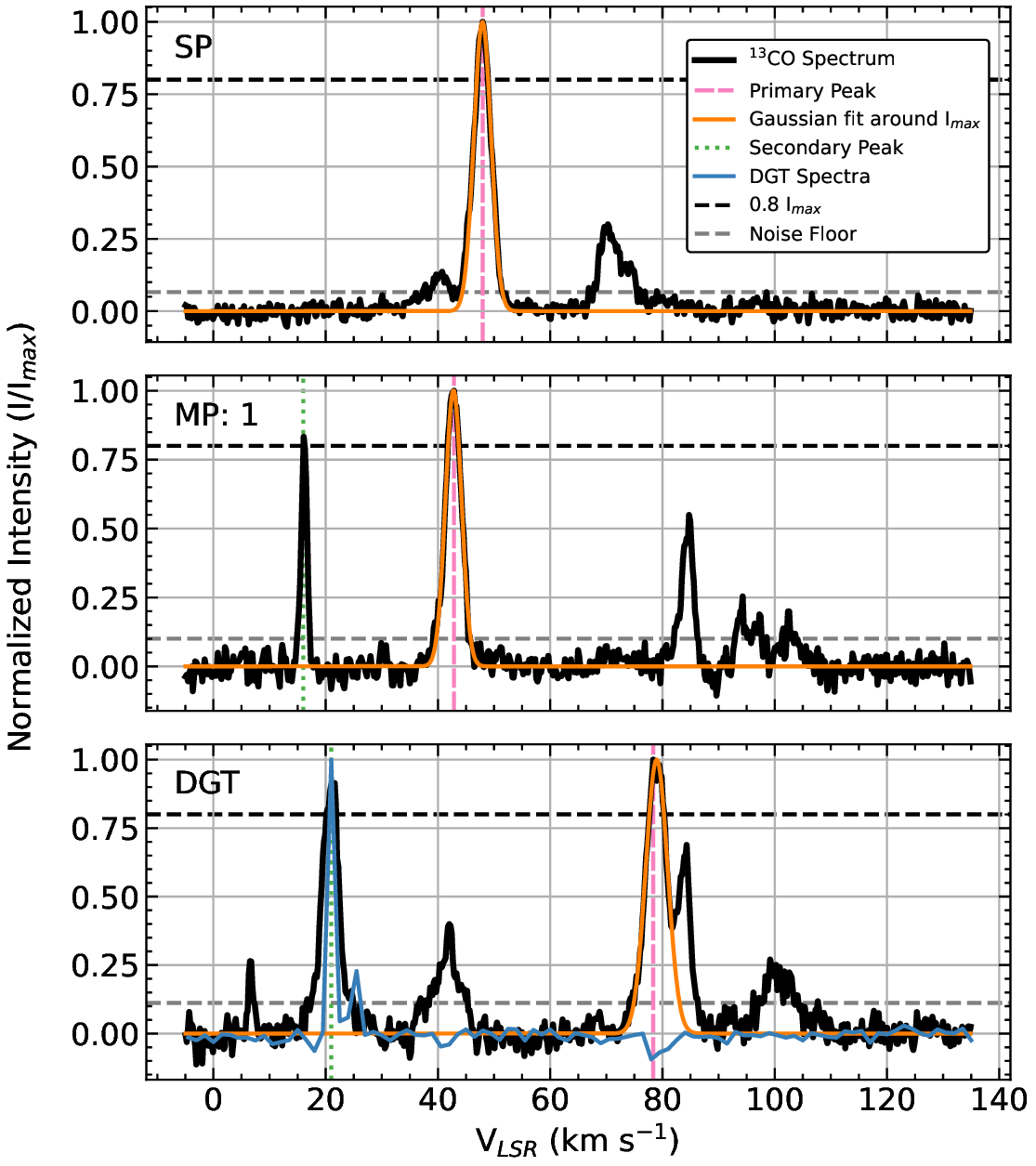}
    \caption{Example MIRION sources showing the different flags used for spectral classification. The top panel shows a SP source, where only one CO peak exceeds 0.8$I_{max}$. The middle panel shows a case where a secondary peak exceeds 0.8$I_{max}$, marking it as a MP: 1 source. Lastly, the bottom panel shows a source associated with a dense gas tracer (DGT). In this case, the secondary peak coincides with a 1662 MHz OH maser from THOR, and is preferentially adopted over the CO peak with the most intense emission.}
    \label{fig:YBSpectra}
\end{figure}

We flagged sources based on the complexity of their spectra as either single-peaked (SP), multi-peaked (MP), low signal-to-noise ratio (LSN), or outside of the CO surveys (OUT). The LSN flag is used to indicate sources with low SNR ($<2$). The MP flag indicates spectra with secondary peaks exceeding 0.8I$\textsubscript{max}$. In this case, we recorded the number of peaks that met this criterion, and flagged the source for further investigation. Although the most intense peak usually corresponds to the most dense gas, it may not trace where star formation is actively occurring. An example of this is shown in \autoref{fig:YBSpectra}. 

To discern whether the CO emission from a MIRION source corresponds to I$\textsubscript{max}$ or a secondary peak, we used spectra from various transitions that trace star formation. We used a SNR of 1.75 for defining peaks in dense gas tracers (DGTs). We identified potential DGT matches using the criterion that the FWHM of the original CO peak (or one of the secondary peaks) overlapped the FWHM of the DGT peak. We then inspected these sources to identify true overlaps.

Because the GRS and SEDIGISM surveys overlap in the range $l$=0.0-18.0$^\circ$, final velocities need to be determined for sources with velocity estimates from both surveys. In the case of a velocity ambiguity, we adopted the GRS velocity over the SEDIGISM, since GRS has superior spectral resolution. In the case of ambiguity with a SEDIGISM velocity outside of the GRS velocity range (i.e. $< -5$ km s$^{-1}$), we compared spectral flags (in order of preference, ``SP", ``MP", then ``LSN"). In the case where their spectral flags match, we compared their integrated distance probabilities from Reid (see \autoref{sec:distance_results}). If no CO velocity was found using the available surveys, but the MIRION source spatially matched within $24''$ of a Hi-GAL compact source with a known distance \citep{Mege2021} (hereafter M2021), we adopted the corresponding velocity. 

Velocity estimates for all 6176 MIRION sources were obtained using the above methods. Of the primary surveys used, 1830 (29.5$\%$) were determined from the GRS, 2612 (42.3$\%$) from SEDIGISM, 240 (3.9$\%$) from FCRAO SMOG, 597 (9.7$\%$) from ThrUMMS, and 570 (9.2$\%$) from low-resolution CO surveys. The remaining 327 (5.3$\%$) of sources have velocities taken directly from M2021, due to lack of survey coverage or because they met the non-kinematic M2021 criteria. In terms of their spectral flags, 4085 (66.1$\%$) are SP, 849 (13.8$\%$) are LSN, 416 (6.7$\%$) are OUT, and 826 (13.4$\%$) are MP. Of the sources that are MP, 651 contained a single additional peak (MP:1), 127 contained two (MP:2), and 48 contained more than two additional peaks. It is important to note that most of the sources with more than two additional peaks arise from SEDIGISM spectra which approach LSN.  Of the MP and LSN sources, 107 (2$\%$) have DGT velocities that differ from the original CO estimates. 

%high-reliability: those with three adjacent fluxes in the spectral range $160\mu m \leq \lambda \leq 500 \mu m$, a concave down shape, and $F_\text{350}-F_{500} > 0$ \citep{Elia2021})

\subsection{Distance Results}
\label{sec:distance_results}
With position and velocity information available for each MIRION source, distances can be estimated using a Galactic rotation model. In our case, we used the Parallax-Based Distance Calculator V2 from \citet{Reid2019}. This distance calculator takes as inputs the MIRION source (l,b) coordinates, the LSR velocity and the $P\textsubscript{far}$ parameter, which is the probability that the object lies at the far kinematic distance. The \citet{Reid2019} distance calculator provides a probability-based resolution of the near-far distance ambiguity resulting from using velocity information to estimate distance in the inner galaxy.

To assign $P\textsubscript{far}$ to the MIRION objects, we obtained Hi-GAL distance results from M2021. M2021 used a variety of distance determination methods to disambiguate Hi-GAL compact sources. Using their results, we compared near ($d\textsubscript{near}$), far ($d\textsubscript{far}$), and adopted distances ($d$) for spatially-matched MIRION sources (priority order 3-9 in table 2 from M2021). If, for instance, $d\textsubscript{near} = d$, we set the $P\textsubscript{far}$ value to 0.0. A status of MDIST indicates the source distance was determined using maser parallax, while a status of SDIST$\_$GROUP means the stellar distance was determined from grouping using optical \HII regions. These methods are non-kinematic and subject to less ambiguity; therefore, we adopted their distances directly. For sources with NO$\_$KDA (no solution at a given velocity and therefore far distance adopted), KDA$\_$NO (no velocity available),
%hence the velocity is set to -999) 
TGT$\_$POINT (tangent point distance, or the distance adopted at the highest allowed velocity for the model at a given (l,b) when the CO velocity exceeded that value), or those lacking a status outside of the M2021 catalog, we used a constant $P\textsubscript{far}$=0.5. 

With this information, the Reid calculator outputs the estimated distance, error, total integrated probability of the source being at that distance, and in which spiral arm the source is located. Following the same method as with the velocity determination, if no velocity (and therefore no distance) was found from the CO surveys, but the MIRION source matched a Hi-GAL compact source with a known distance (M2021), we adopted that distance. We determined distance estimates for $\sim 94\%$ of MIRION sources.

\begin{deluxetable}{lc}
\tablecaption{Acronyms \label{tab:Surveys}}
\tablewidth{0pt}
\tabletypesize{\normalsize}
\tablehead{\colhead{Acronym}  & \colhead{Description}}
\startdata
SP              & Single peaked CO spectrum \\
MP              & Multi-peaked CO spectrum \\ 
LSN             & Low signal-to-noise ratio \\ 
OUT             & Velocity adopted from \citet{Mege2021} \\ 
MDIST*          & Maser Distance \\
SDIST$\_$GROUP* & Stellar distance from grouping \\
KDA$\_$NO*      & No velocity \\
NO$\_$KDA*      & No solution \\
TGT$\_$POINT*   & Tangent point distance (forbidden velocity)  \\
NO$\_$AMB*      & No kinematic ambiguity \\
\enddata
\tablecomments{Acronyms presented in this paper. Those with an asterisk are from from \citet{Mege2021} table 2.} 
\end{deluxetable}

%\begin{figure}
    %\centering    %\includegraphics[width=0.75\linewidth]{Figures/YBs_top_down_figure.eps}
    %\caption{Top down plot showing the locations of all 6176 MIRION sources. The primary surveys and regions are marked with colored dashed lines. The green ``TGT PT" indicates the sources that have tangent point distances: these may be non-physical, and are  due to measured velocities exceeding the maximum allowed velocity for the rotation model at a given (l,b).}
    %\label{fig:TopDown}
%\end{figure}

\subsection{Comparison with Previous Studies}

% NL: Velocity comparison and distance comparison split into two paragraphs
% Break after "10 km/s respectively"
% Lead with "mean distance difference" go into what fraction
% Something like "we conclude.... of comprable accuracy..... other kinetmatic based tech." 
% Not purely kinematic based? Watch wording on above 

We selected a set of MIRION sources that were also M2021 matches to compare our results with previous studies. Only sources with independent distance estimates were included, totaling 2666 objects. M2021 compared their radial velocity measurements for 5976 sources with those from ATLASGAL \citep{2018Urquhart}, and found that 89.4$\%$ and 91.8$\%$ of their sources had velocity differences smaller than 5 km s$^{-1}$ and 10 km s$^{-1}$, respectively. Comparing MIRION velocities to M2021, we found that 80.2$\%$ and 83.3$\%$ of matched sources had velocity differences smaller than 5 km s$^{-1}$ and 10 km s$^{-1}$, respectively. 

Additionally, M2021 compared their distances to ATLASGAL by selecting sources outside $\pm$12$^\circ$ of the Galactic Center (due to highly uncertain distance determination), including only those with a velocity difference less than 5 km s$^{-1}$, and not flagged as ``NO$\_$KDA". They found a mean distance difference of 0.21 kpc, with 2248 (64$\%$) of their sources having a difference $<0.7$ kpc. When we compare independent MIRION distances with those of M2021 meeting the same criteria, we find a mean distance difference of $0.3$ kpc, with 61$\%$ of 1973 matched sources having differences less than 0.7 kpc. 

\section{Source Crossmatching} \label{sec:xmatch}
To identify MIRION source associations in other catalogs, we crossmatched sources to the same catalogs used in WKD21. We retrieved these catalogs using the VizieR catalog access tool \citep{vizier}. We performed a $24''$ crossmatch with the Hi-GAL compact source catalog II (3979 sources, \citealt{Elia2021}), the ATLASGAL dust condensation catalog (1623 sources, \citealt{Csengeri2014}), the ATLASGAL compact source catalog (CSC) (1563 sources, \citealt{Contreras2013}), the RMS survey (613 sources, \citealt{Lumsden2013}), the WISE survey (2315 total sources, \citealt{Anderson2014}), as well as the CORNISH North (192 sources, \citealt{Purcell2013ApJ}) and South (384 sources, \citealt{Irabor2023}) surveys. The results of this are presented in \autoref{tab:CrossMatch}.  

The Hi-GAL and both ATLASGAL surveys identify compact sources (i.e. clumps, cores) at various stages of evolution using far-IR and sub-mm observations. The ATLASGAL GaussClump Source Catalog (GCSC, \citet{2014Csengeri}) used the GaussClump algorithm, which assumes a Gaussian intensity distribution and is sensitive to angular scales corresponding to both nearby cores and distant clumps. The ATLASGAL CSC used Source-Extractor, which is better suited for molecular clump and cloud structures on a larger scale than the GCSC. 

The RMS survey searched for young massive stellar objects and used supplemental surveys at other wavelengths to classify different types of objects (e.g. YSOs, \ion{H}{2} regions, Planetary Nebulae). The WISE survey used MIR emission and source morphology to identify candidate and known \ion{H}{2} regions, distinguishing sources into categories ``K" for known \ion{H}{2} regions, ``C" for candidate \ion{H}{2} regions, ``G" for group, and ``Q" for radio-quiet sources. The CORNISH surveys used 5 GHz radio continuum to identify \ion{H}{2} regions.

To ensure that the crossmatches were not spurious in nature, we created 100 simulated distributions of MIRION sources for each survey. We selected random Galactic longitudes from a uniform distribution within the same longitude range as the surveys. We selected latitudes for each source from Gaussian distributions parameterized by the MIRION source extent in Galactic latitude;  for the first and fourth Galactic quadrant we used $\mu$ = -0.06$^\circ$, $\sigma$ = 0.42$^\circ$, and for Cygnus X we used $\mu$ = 0.73$^\circ$, $\sigma$ = 1.28$^\circ$. The MIRION sources from the SMOG region are offset from the Galactic mid-plane and do not resemble a Gaussian distribution. In order to better estimate the Galactic latitude distribution, RMS \ion{H}{2} regions / YSOs were used as a proxy for MIRION sources, which (over the SMOG region) exhibit a Gaussian distribution with $\mu$ = 0.61$^\circ$ and $\sigma$ = 1.74$^\circ$.

% GWC Question - Distributions should be Gaussian with peak near the mid-plane (GLAT = 0). I assume this was taken into account for Cyg-X and SMOG, which are above the mid-plane?
% CK - look at defining the mid-plane not using the catalogs
% NL answer - updated from work done October, 2024. Uniform distirbution un-physical, does not resemble a Gaussian. See email chain 10/18/2024, and notes in Nicks Notes / Fake YB Cross Matching Update on Google Drive. (mean and std here are for HII regions / YSOs, whereas those in Google Drive notes were originally for 8um sources). 
 
The simulated MIRION sources were crossmatched to each catalog. This resulted in matches with $3.4\% \pm 0.2\%$ (210 $\pm$ 13) of the randomly generated MIRION sources for the Hi-GAL catalog (within the HI-GAL survey region). For all other surveys, less than $1\%$ of the generated MIRION sources matched with the star formation catalogs. These results are consistent with those presented in WKD21, where the Hi-GAL catalog matched with $4.5\% \pm 0.9\%$ of simulated sources. We maintain that catalog crossmatches can reliably be used to associate MIRION sourcs with tracers of star formation given the low level of spurious association. 

The source ``hit rate" reported by the MWP in the creation of the original YB catalog (see Table~\ref{tab:mirion-phot}) is another metric by which to judge whether MIRION sources are truly associated with sites of star formation. The hit rate for MIRION sources is the ratio of the total number of times a source was identified by MWP users to the total number of times images containing the source were viewed (WKD21). The hit rate for all 6176 MIRION sources shows a wide distribution, with a mean and standard deviation of 0.40 and 0.19 respectively. This is similar to the distribution of the sample WKD21 studied. The matched surveys and their corresponding mean hit rates for MIRION sources are Hi-GAL (0.43), ATLASGAL Condensations (0.43), ATLASGAL compact sources (0.44), RMS (0.42), WISE (0.48), CORNISH N (0.46), and CORNISH S (0.46), all with  standard deviation $\sim0.20$. The mean hit rate for the 1457 MIRION sources without any catalog crossmatches is 0.32 with a standard deviation of 0.15. The difference between the mean hit rates of MIRION sources with and without catalog matches is very similar to what WKD21 found. The lower mean hit rate for unmatched MIRION sources may be explained in part by the fact that many of the unmatched sources have low fluxes and would have been missed by these surveys, although some unmatched objects may be ISM features identified by MWP volunteers that are not actually associated with star-forming regions. %[GWC - Do we want to include Nick's flux CDF figures here?]

%CK Commented out most of the subsets - can easily put them back in if we decide we prefer to have that information too.
\begin{deluxetable}{lccc}
\label{tab:CrossMatch}
\tablecaption{MIRION Source Crossmatch Results \label{tab:CrossMatch}}
\tablewidth{0pt}
\tabletypesize{\normalsize}
\tablehead{\colhead{Survey}  & \colhead{Subset} & \colhead{Num. Matches}       & \colhead{Percent Match (Survey Range)} } 
\startdata
Hi-GAL CSC-360         &                & 3979  & 65.26\%   \\
ATLASGAL Condensations &                & 1623  & 30.48\%   \\
ATLASGAL CSC2014       &                & 1563  & 41.24\%   \\
RED MSX                &           & 613   & 11.03\%   \\
%RED MSX                & Carbon Star    & 1     & 0.02\%    \\
%RED MSX                & Evolved Star   & 10    & 0.18\%    \\
%RED MSX                & OH/IR Star     & 6     & 0.11\%    \\
%RED MSX                & Young/Old Star & 2     & 0.04\%    \\
%RED MSX                & YSO            & 109   & 1.96\%    \\
%RED MSX                & PN             & 10    & 0.18\%    \\
%RED MSX                & Proto-PN       & 0     & 0.00\%    \\
%RED MSX                & \ion{H}{2} Region    & 455   & 8.19\%    \\
%RED MSX                & \HII/YSO       & 16    & 0.29\%    \\
%RED MSX                & Other          & 4     & 0.07\%    \\
%RED MSX                & M              & 26    & 0.47\%    \\
WISE                   & Total          & 2315  & 37.48\%   \\
WISE                   & K              & 435   & 7.04\%    \\
WISE                   & G              & 113   & 1.83\%    \\
WISE                   & C              & 486   & 7.87\%    \\
WISE                   & Q              & 1276  & 20.66\%   \\
%WISE                   & ?              & 5     & 0.08\%    \\
CORNISH                &           & 576   & 12.08\%   \\
%CORNISH South          &                & 384   & 15.42\%   \\
%CORNISH North          &                & 192   & 8.44\%                                 
\enddata
\tablecomments{Results from the 24$''$ crossmatch of the MIRION sources with catalogs tracing star formation. Column 4 indicates the fraction of MIRION sources that match with a given object out of the total number of MIRION sources that exist within that survey range.}
\end{deluxetable}

\section{Analysis} \label{sec:analysis}

\subsection{Physical Properties of Hi-GAL Associated Clumps} \label{sec:clumps}

\begin{figure}
\plotone{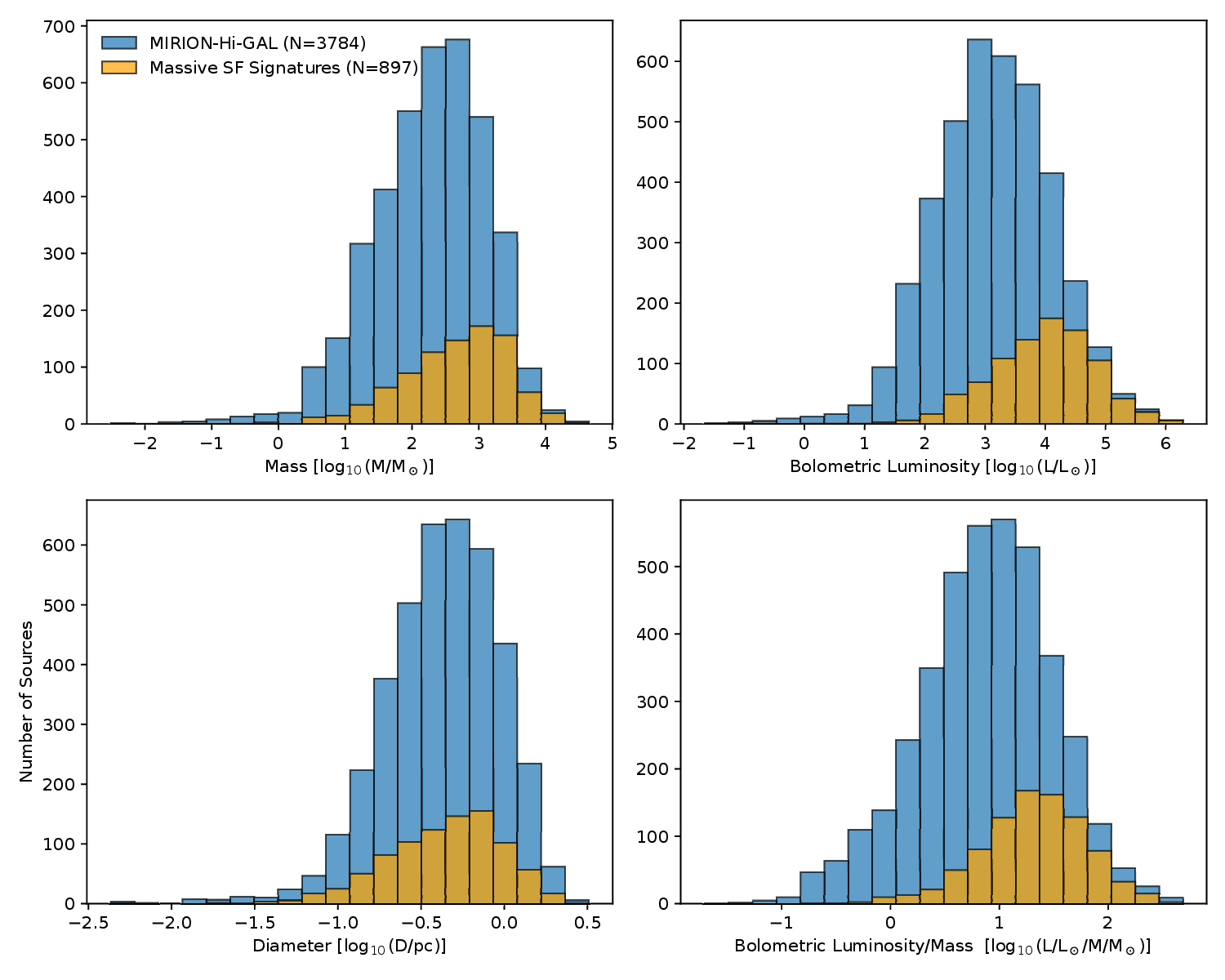}
\caption{Physical properties of MIRION catalog--Hi-GAL matched sources. Lower-left and upper-row histograms show values from \citet{Elia2021} rescaled using our newly calculated distances. The lower-right panel shows the distance-independent luminosity-mass ratio from \citet{Elia2021}. Smaller histograms show the distribution of properties for MIRION objects with clear signatures of massive star formation. See text for full discussion.}
\label{fig:histo1}
\end{figure}

\begin{figure}
\plotone{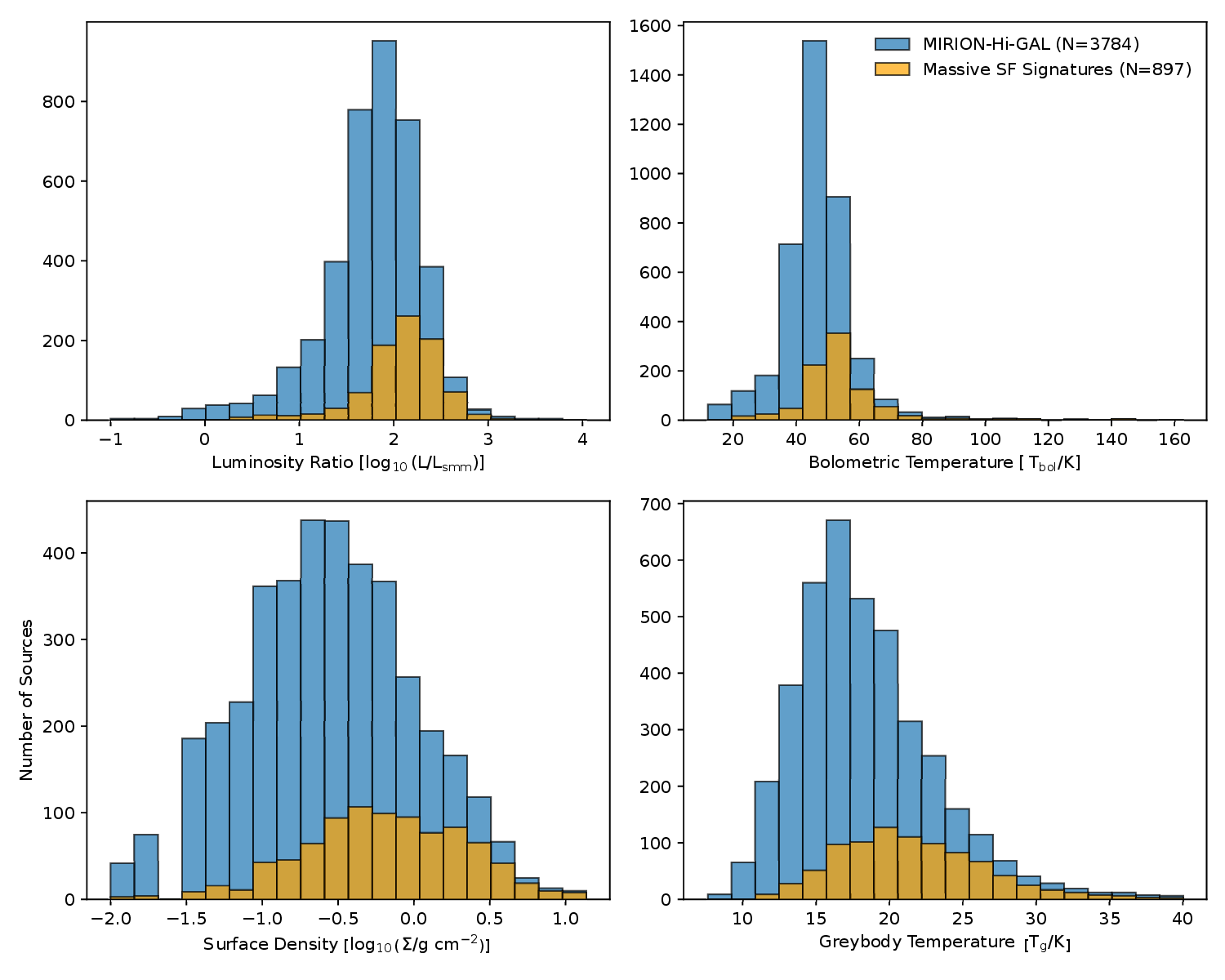}
\caption{As Figure~\ref{fig:histo1}. All quantities plotted are distance-independent.}
\label{fig:histo2}
\end{figure}

The subset of MIRION sources with Hi-GAL CSC associations are particularly interesting as they are the objects that are most likely associated with star-formation activity. As described in Section~\ref{sec:catalog}, Table~\ref{tab:mirion-hcsc} lists these objects along with both distance-independent and rescaled distance-dependent physical properties. Histograms showing the distribution of the physical properties are shown in Figures~\ref{fig:histo1} and \ref{fig:histo2}, and descriptive statistics are listed in the top rows of Table~\ref{tab:stats}. For this sample we removed 34 sources that were missing one or more physical properties, resulting in a sample of 3945 MIRION sources.   

Focusing first on distance-dependent quantities, we find that most of these MIRION sources have masses in the range $0 < \log_{10}(M/\mathrm{M_\sun}) < 4$, luminosities in the range $1 < \log_{10}(L/\mathrm{L_\sun}) < 5$, and diameters falling in the canonical clump range of 0.2 -- 3 pc. These results are similar to those found by WKD21 (compare Table~\ref{tab:stats} with Table 5 in WKD21). 

Distance-independent quantities can be related to the evolutionary stage of the clump or to the mass of the stellar population forming within a clump. Large $L/M$ ratio and surface density values are typically associated with massive SFRs \citep[e.g.,][]{Elia2017}. The luminosity ratio, $T_\mathrm{bol}$ and, to a lesser extent, $T_\mathrm{gray}$ are all evolutionary stage proxies. The increase in luminosity ratio and $T_\mathrm{bol}$ reflect the expected shift in the clump SED peak to shorter wavelengths at later evolutionary stages, and the increase in $T_\mathrm{gray}$ reflects the overall ISM temperature increase expected to occur as stars form within a clump. Comparing Table~\ref{tab:stats} with Table 5 from WKD21 we can observe that the histogram properties are again very similar. Table~\ref{tab:thresholds} summarizes some key threshold values for four of these quantities and compares our results with those found by WKD21, and values derived for the full protostellar Herschel clump sample from \citet{Elia2017}.

%CK - Not sure if we really need the paragrph below now. Essentially I'm trying to show that the MIRION sample is very similar on average to the WKD21 sample - perhaps that's clear enough from the table and from the explicit statement at the end of the second paragraph of this section.
%
%We see that our MIRION sample contains a slightly lower percentage of high-mass or highly evolved regions than the WKD21 sample, but  a higher percentage of such regions than the full Elia 2017 protostellar clump sample. In short, our larger MIRION sample solidifies the conclusions of WKD21, 
%Note to CK - If these conclusions are referenced here, it should also be stated to which conclusions you are actually refering.
%and clearly demonstrates that the MIRION sources with Hi-GAL CSC associations are SFRs covering a wide range of masses and evolutionary stages. 

It is interesting to explore how clump physical properties vary based on different subsets selected via the catalog crossmatching described in Section~\ref{sec:xmatch}. One simple division is to divide the sample into MIRION sources with and without clear signatures of high-mass star formation. A high-mass signature subsample of 897 objects was defined as MIRION sources having WISE types of C, K, or G; RMS types of \ion{H}{2} region or \ion{H}{2}/YSO; or CORNISH types of UC\ion{H}{2}, \ion{H}{2} region (including Diffuse and Dark), MYSO, or IR Quiet. Objects with multiple catalog crossmatches had to meet these criteria in all catalogs. For example, a MIRION source with CORNISH \ion{H}{2} region and WISE~K crossmatches would be included, but a source with CORNISH \ion{H}{2} region and WISE~Q crossmatches would not. The distribution of physical properties for this subset is shown as the gold histograms in Figures~\ref{fig:histo1} and \ref{fig:histo2}, and descriptive statistics are listed in the middle rows of Table~\ref{tab:stats}. Descriptive statistics for the complementary sample of objects with no high-mass star-formation signatures are shown in the lower rows of Table~\ref{tab:stats}. For this sample we have removed the stellar and planetary nebula crossmatches resulting in a sample of 2887 sources. 

As expected, the histograms associated with signatures of high-mass star formation are skewed towards higher masses and luminosities (top row of Figure~\ref{fig:histo1}). There is no significant difference in the diameter distributions of the high-mass sample compared to the full sample. All of the histograms of distance-independent quantities are shifted to larger values in the high-mass sample, as expected for sources that are involved in high-mass star formation, more evolved, or both. Interestingly, while the average surface density also shifts to higher values for this sample, the range of these values remains very large. This is likely the result of high-mass SFRs in our sample being at different stages of evolution. For example, MIRION sources with either an UC\ion{H}{2} region association or a Diffuse \ion{H}{2} association would both be included, but the associated surface densities are likely to be vastly different due to the clearing of material expected at later stages of massive SFR evolution.

\begin{figure}
    \centering
    \includegraphics[width=0.5\textwidth]{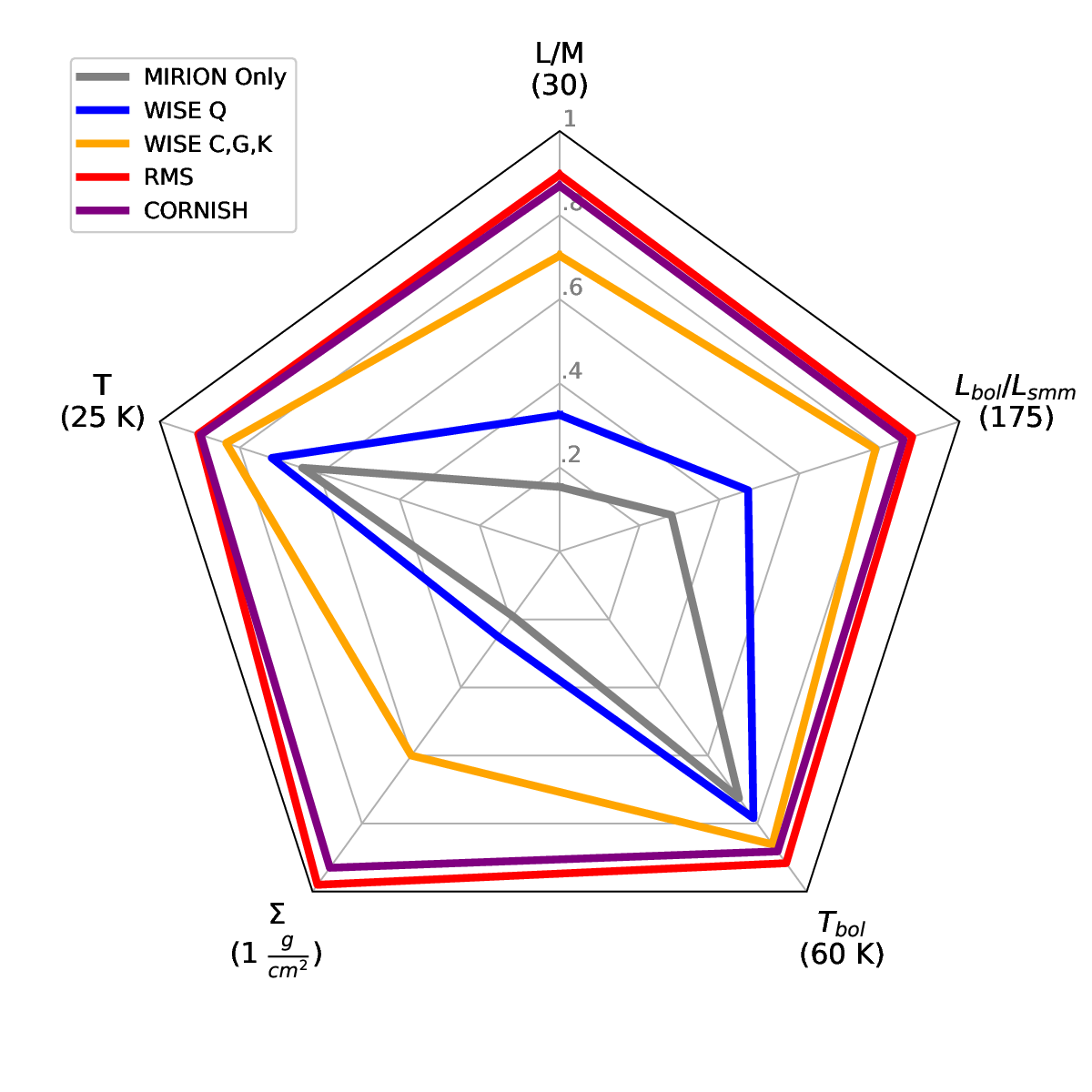}
    \caption{Radar plot representation of five distance-independent quantities for subsets of MIRION sources with Hi-GAL crossmatches. The values shown are medians and are plotted as fractions of the maxima indicated at each radial line. Subsets are based off of source crossmatches with other catalogs, indicated in the legend. $L/M$: Bolometric luminosity to clump mass ratio; $L_{bol}/L_{smm}$: Ratio of bolometric luminosity to luminosity calculated in the submillimeter range; $T_{bol}$: Bolometric temperature; $\Sigma$: Clump surface density; $T$: Dust temperature.}
    \label{fig:radarplot}
\end{figure}

The radar plot shown in Figure~\ref{fig:radarplot} compares median distance-independent values for five different crossmatch subsets. We see the CORNISH and RMS subsets are essentially identical in this five-parameter space and have median values expected for evolved, high-mass SFRs. The WISE C/G/K subsample shows there is a significant decrease in the median surface density. This most likely reflects the fact that the WISE-based sample includes more evolved \ion{H}{2} regions, which will tend to have a lower surface density due to the removal and dispersal of material caused by the combined action of ionization and stellar winds.

The WISE Q and MIRION only subsamples have median properties that clearly differ from the other subsets to a significant extent. It is tempting to simply conclude that this difference reflects the inability of these clumps to form high-mass stars and that WISE Q and MIRION only clumps are mostly all low- and intermediate-mass SFRs. Such a conclusion, however, fails to consider the impact of these objects' evolutionary stage of their properties.

\begin{figure}
\plotone{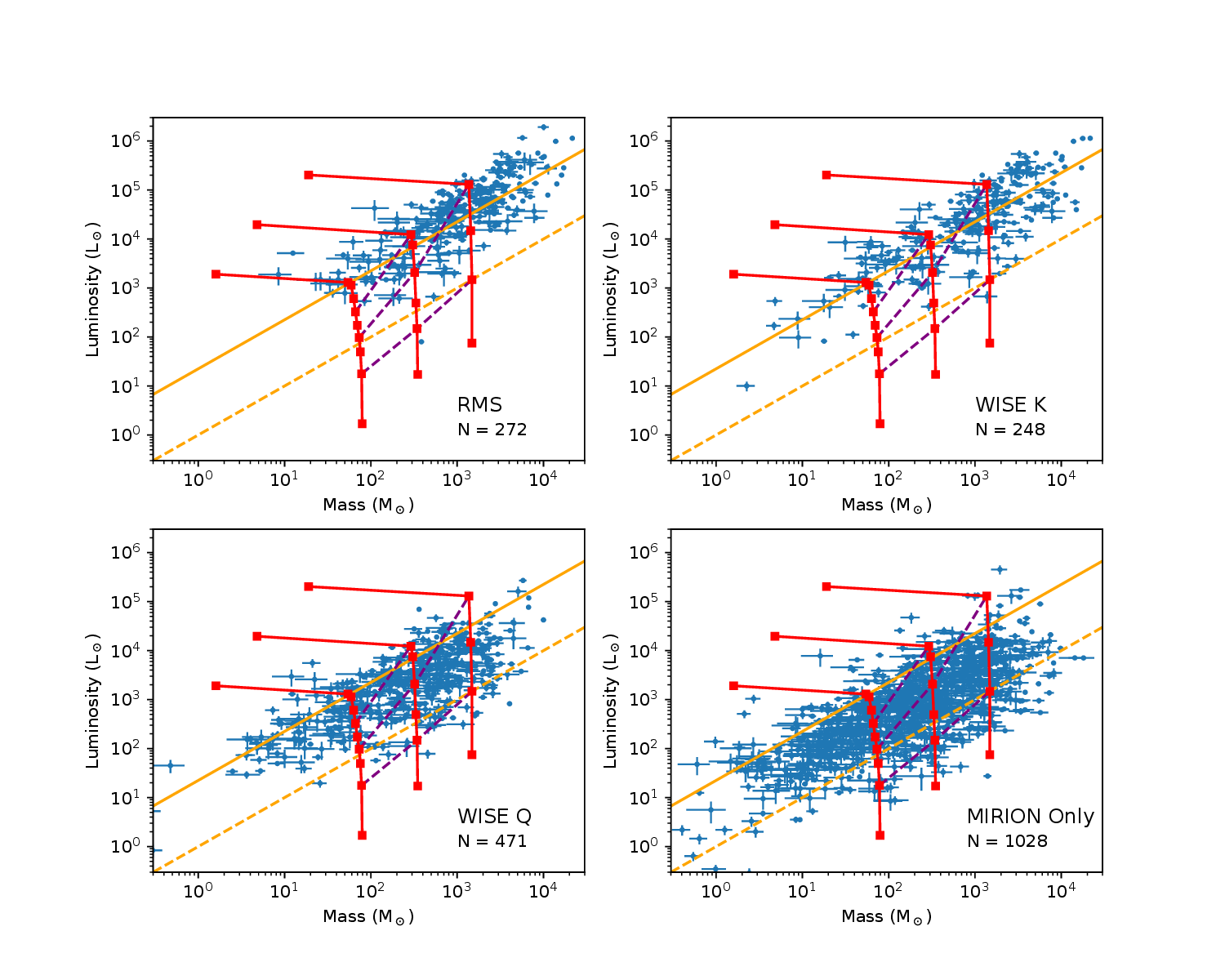}
\caption{Luminosity-Mass plots for MIRION Catalog sources with Herschel CSC matches. Only sources with fractional error in mass $< 0.5$ are plotted. Each panel corresponds to objects with a particular catalog association as shown. The yellow diagonal lines correspond to the 90th percentile lower-limit for Herschel-defined \HII regions (solid) and the 90th percentile upper-limit for pre-stellar sources (dashed) both defined in \citet{Elia2017}. \citet{Mol2008} clump evolutionary tracks for 80, 350 and 2000 \msol clumps are shown (red solid lines). Red squares indicate $5\times10^4$ year intervals on the vertical portions of each track. Three representative isochrones, separated by $10^5$ years, are shown as dashed purple lines. 
}
\label{fig:LMebar}
\end{figure}

The effect of clump evolution on observed clump properties is best illustrated using a luminosity-mass (LM) plot. Figure~\ref{fig:LMebar} shows LM plots for four different subsets of MIRION sources. All of the sources selected have low fractional mass and luminosity uncertainties ($<0.5$).

The upper-row plots show sources with high-mass star-formation signatures from the RMS and WISE-K surveys. The majority of these points lie above the 90th percentile lower-limit for \ion{H}{2} region associated clumps defined by \citet{Elia2017}. The lower-row plots show sources that do not have high-mass star-formation signatures. As expected for clumps having some star-formation activity, the majority of these sources lie above the 90th percentile upper-limit for pre-stellar clumps defined by \citet{Elia2017}. Comparing this figure to Figure 4 in WKD21, we see the full survey populates the 10 -- 100 \msol region more fully, with 378 sources ($\sim 25\%$ of the WISE~Q and MIRION Only combined sample) falling in this mass range.

In each plot we have overplotted representative evolutionary tracks and isochrones from the \citet{Mol2008} model for clump evolution. In this model the star-formation process starts in isolated pre-stellar clumps having a particular mass. The clump then evolves almost vertically on the LM plot as a single intermediate- or high-mass star forms within the clump, reducing its mass slightly while raising the luminosity. The star-formation stage then ends with a dispersal stage in which the luminosity of the clump remains roughly constant as the clump mass decreases. In this model, the observed mass of the clump constrains its future evolutionary path. For example, any clump with a mass between 80 -- 350 \msol should follow an evolutionary path  resulting in a SFR with a (stellar) luminosity between $2\times10^3 - 2\times10^4$ \lsol ($\log_{10}(L/\mathrm{L}_\odot)=3.3-4.3$). This luminosity range is at the dividing line between high-mass (OB) stars and lower-mass later B-type stars, meaning we can use evolutionary tracks in the 80 -- 350 \msol range as a dividing line between clumps associated with high-mass and intermediate/low-mass star formation. 

The RMS and WISE K samples in Figure~\ref{fig:LMebar} are, by definition, high-mass SFRs, and we find that only 24\% of the combined sample have a clump mass $<350$~\msol. In contrast, 57\% of the combined WISE Q and MIRION-only sample have clump masses $<350$~\msol. In the context of the \citet{Mol2008} evolutionary models, these sources are likely IMSFRs. If we remove the fractional error constraint on these samples this percentage increase to 66\%. 

We note that more recent models of star formation in clumps \citep{Molinari2019} also model the formation of multiple stars within a clump, but they do not try to account for the fact that clumps are not observed to be fully isolated from their surrounding environment and that star formation in clumps likely involves the flow and accretion of material at multiple scales as star formation occurs \citep{Motte2018}. As a result, the vertical, star-formation portion of the evolutionary tracks shown in Figure~\ref{fig:LMebar} become tilted to the right, reflecting the fact that clump mass can increase as star formation occurs. Unfortunately, the evolutionary tracks in this picture are more complex and can become degenerate with the evolutionary track for any particular clump becoming highly dependent on the clump's surrounding environment (\citealt{KM2020}, Larose et al.\ in preparation). The range of masses and luminosities spanned by the MIRION catalog makes it an ideal dataset for testing various clump evolutionary models.

\begin{deluxetable}{lccccccccc}
\tabletypesize{\small}
\tablecaption{Descriptive Statistics for Hi-GAL Matched Sources \label{tab:stats}}
\tablewidth{0pt}
\tablehead{\colhead{} & \colhead{Mass} & \colhead{Luminosity} & \colhead{Diameter} & \colhead{ L$_\mathrm{bol}$/Mass} & \colhead{T$_\mathrm{grey}$} & \colhead{L$_\mathrm{ratio}$} & \colhead{T$_\mathrm{bol}$} & \colhead{Surface Density} \\
\colhead{} & \colhead{$\log_{10}\left(\mathrm{M_\sun}\right)$} & \colhead{$\log_{10}\left(\mathrm{L_\sun}\right)$} & \colhead{$\log_{10}\left(\mathrm{pc}\right)$} &  \colhead{$\log_{10}\left(\mathrm {L_\sun/M_\sun }\right)$} & \colhead{(K)} & \colhead{$\log_{10}\left(L/L_\mathrm{smm}\right)$} & \colhead{(K)} & \colhead{$\log_{10}\left(\mathrm{g~cm}^{-2}\right)$}}
\startdata
Median & 2.34 & 3.14 & $-0.35$ & 0.91 & 17.6 & 1.85 & 46.5 & $-0.57$ \\
Mean   & 2.26 & 3.13 & $-0.38$ & 0.87 & 18.3 & 1.78 & 46.8 & $-0.56$ \\
SD     & 0.85    & 0.99    & 0.35    & 0.62    & 4.7  & 0.54    & 11.9  & 0.56 \\
Max.   & 4.65    & 6.28    & 0.51    & 2.68    & 40.0 & 4.04    & 162.7 & 1.14 \\
Min.   & $-2.51$ & $-1.66$ & $-2.37$ & $-1.71$ & 7.6  & $-1.00$ & 11.8  & $-2.00$ \\
\hline
Median & 2.78     & 4.03 & $-0.31$ & 1.32    & 21.2 & 2.13    & 52.0  & $-0.19$   \\
Mean   & 2.65     & 3.95 & $-0.37$ & 1.30    & 21.6 & 2.07    & 53.5  & $-0.20$   \\ 
SD     & 0.79     & 0.86 & 0.35    & 0.49    & 5.1  & 0.46    & 13.7  & 0.53    \\ 
Max.   & 4.58     & 6.28 & 0.51    & 2.67    & 39.1 & 3.49    & 146.4 & 1.14    \\
Min.   & $-1.25$  & 0.10 & $-2.29$ & $-0.68$ & 8.5  & $-0.52$ & 17.5  & $-2.00$ \\
\hline
Median & 2.21    & 2.89    &  $-0.36$ & 0.75    &  16.6 & 1.75    & 44.7  & $-0.70$ \\
Mean   & 2.13    & 2.84    &  $-0.38$ & 0.71    &  17.2 & 1.67    & 44.2  & $-0.69$ \\
SD     & 0.83    & 0.87    &  0.35    & 0.58    &  4.1  & 0.52    & 10.1  & 0.51    \\
Max.   & 4.65    & 5.65    &  0.42    & 2.68    &  40.0 & 4.04    & 162.7 & 1.13    \\
Min.   & $-2.51$ & $-1.66$ &  $-2.37$ & $-1.71$ &  7.6  & $-1.00$ & 11.8  & $-2.00$ \\
\enddata
\tablecomments{Statistics for three different samples are shown. Top: full MIRION--Hi-GAL sample (3945 sources; large (blue) histograms in Figures~\ref{fig:histo1} and \ref{fig:histo2}). Middle: high-mass SFR signature subsample (897 sources; small (orange) histograms in Figures~\ref{fig:histo1} and \ref{fig:histo2}). Bottom: no high-mass SFR signature subsample (2887 sources). See Section \ref{sec:clumps} for details.}
\end{deluxetable}

\begin{deluxetable}{lcccl}
\tablecaption{Distance Independent Thresholds for Hi-GAL Clumps \label{tab:thresholds}}
\tablewidth{0pt}
\tablehead{\colhead{} & \multicolumn{3}{c}{Percentage of Sample  Meeting Threshold} & \colhead{} \\
\colhead{Threshold} & \colhead{MIRION\tablenotemark{a}} & \colhead{WKD21\tablenotemark{b}} & \colhead{Protostellar\tablenotemark{c}} & \colhead{Threshold Interpretation} 
}
\startdata
$\log_{10}\left(L_{\mathrm{bol}}/M\right) > 1.35$ & 22\% & 24\% & 10\% & Likely high-mass star formation \\
$\log_{10}\left(L/L_\mathrm{smm}\right) \geq 2 $  & 35\% & 40\% & 14\% & Older/evolved SFR \\
$T_{\mathrm bol}<40$ & 20\% & 15\% & 50\% & Younger/new SFR \\
$\log_{10}\left(\Sigma\right)>0$ & 16\% & 21\% & 13\% & Likely high-mass star formation \\
\enddata
\tablenotetext{a}{Hi-GAL matched sample from this paper.}
\tablenotetext{b}{Hi-GAL matched sample from WKD21.}
\tablenotetext{c}{Hi-GAL protostellar clump sample from \citet{Elia2017}}
\end{deluxetable}

\subsection{Photometry-Based Analysis}

As discussed in Section \ref{sec:photom}, we made at least five measurements of each source's flux density at 70, 24, 12, and 8~\um. Multiple measurements control and quantify the uncertainty that results from users' differing selections of the area delineating the sources. The mean flux density and the fractional errors (standard deviation/flux density) at each wavelength are reported in Table \ref{tab:mirion-phot}. There are 4757, 4545, 4176, and 4669 sources at 70, 24, 12, and 8~\um, respectively, having fractional uncertainties less than 0.5. As in WKD21, we use a fractional error of 0.5 as the cut-off for sources' inclusion in the photometry-based analysis.

\begin{figure}
    \centering
    \includegraphics[width=\textwidth]{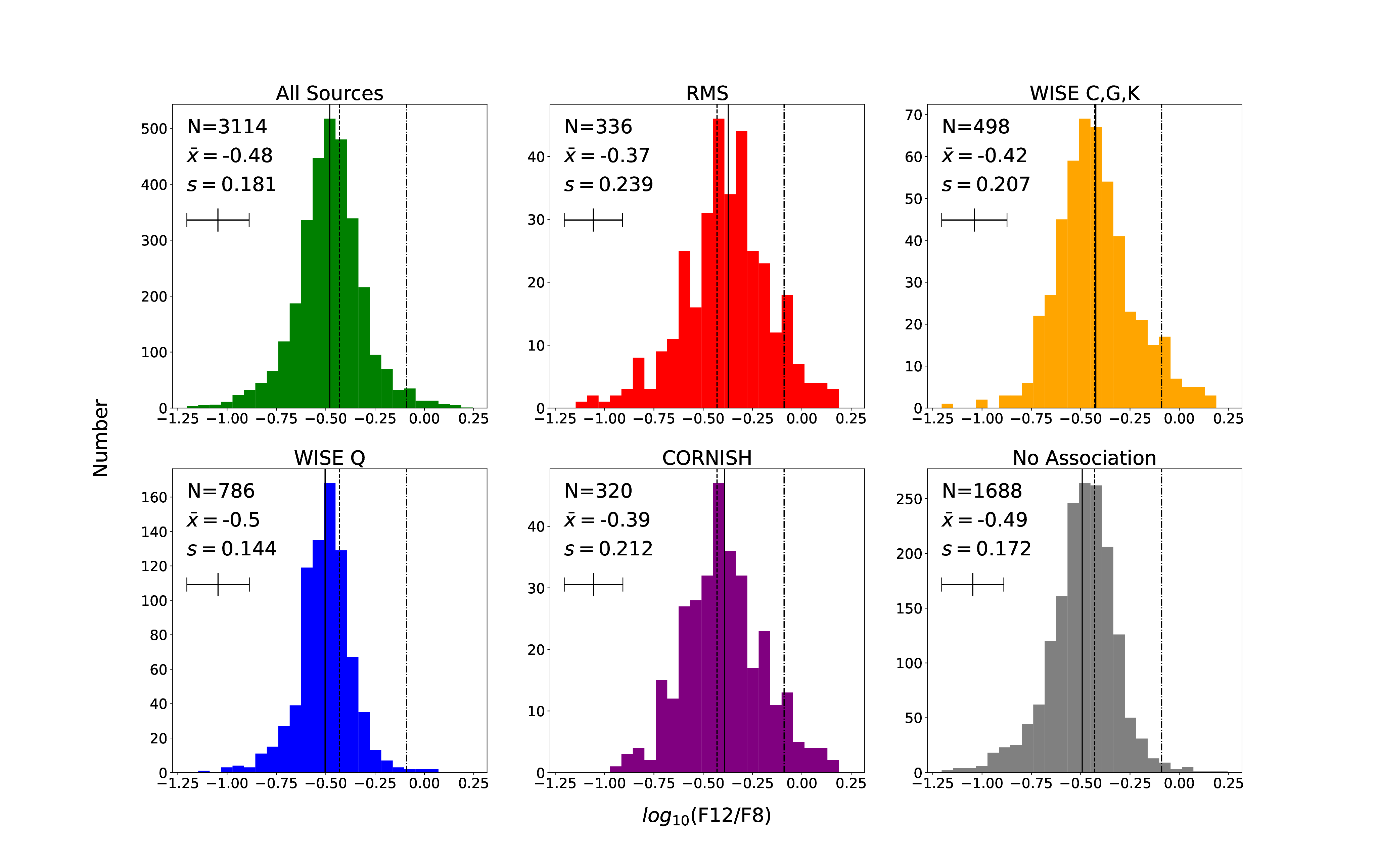}
    \caption{Histograms of the $\log_{10}(F_{12}/F_{8})$ color data collected for MIRION catalog sources. Sources with a fractional error of $>.5$ and sources with Poor Confidence or No Obvious Source flags applied at either 12~\um or 8~\um were excluded from the plots. The sources are categorized by crossmatch, with the top left histogram representing all sources, and the bottom right representing sources with no crossmatch in the RMS, WISE, or CORNISH catalogs. The number of sources ($N$), average color ($\bar{x}$) and standard deviation ($s$) for each plot is shown in the top left corner, as well as a barline representing the $\pm$ average uncertainty in the measurements. The solid line is the average color of sources plotted in the histogram, the dashed line is the average color from the \cite{WC2021} pilot region, and the dot-dashed line is the mean \HII color from \cite{Anderson2012}.}
    \label{fig:colorhist}
\end{figure}

Figure \ref{fig:colorhist} shows the distribution of the number of sources by $\log_{10}(F_{12}/F_{8})$ color, both for the full catalog and for subgroups of sources with crossmatches in other catalogs. Similar histograms were produced in KWA15 and WKD21, but this study increases the number of sources with both $F_{12}$ and $F_{8}$ values below the 0.5 fractional error cut-off by nearly an order of magnitude (3116 sources, as opposed to 324 in WKD21). We find a mean $\log_{10}(F_{12}/F_{8})$ color ($\pm$ standard deviation) of $-0.48 \pm 0.19$ for the full sample, very close to the mean value of $-0.43 \pm 0.15$ reported by WKD21. %Uncertainty is estimated using the average uncertainty for $\log_{10}(F_{12}/F_{8})$ value for the sources included in the analysis.
The solid lines in Figure \ref{fig:colorhist} represent the mean value for each plot, while the dashed vertical lines represent the mean value of $-0.43$ from the WKD21 pilot region.

In their analysis of YB colors, KWA15 and WKD21 explain that a more compact PDR, with a correspondingly denser radiation field, has a lower $\log_{10}(F_{12}/F_{8})$ color. The $\log_{10}(F_{12}/F_{8})$ ratio serves as a measure of compactness and PAH ionization. The average $\log_{10}(F_{12}/F_{8})$ colors for sources with crossmatches in the RMS, WISE C/G/K, and CORNISH catalogs are $-0.37 \pm 0.24$, $-0.42 \pm 0.21$, and $-0.39 \pm 0.21$, respectively. Notably, all of these subgroups have mean $\log_{10}(F_{12}/F_{8})$ colors significantly below the average WISE \HII region color of $-0.09$ reported by \citet{Anderson2012}, indicating that MIRION sources correlating to \HII regions and MYSOs are still more compact than most WISE catalog sources.

Sources with no crossmatches in the WISE, RMS, or CORNISH catalogs (``No Association" in Fig. \ref{fig:colorhist}) have a mean value of $-0.49 \pm 0.18$, and sources crossmatched with the ``radio-quiet" WISE Q sources have a mean value of $-0.51 \pm 0.15$. These values indicate that MIRION sources without \HII region and/or MYSO associations are even more compact. These sources represent $\sim 80 \%$ of the 3114 sources plotted in Figure \ref{fig:colorhist}, further supporting the conclusion from WKD21 that a majority of MIRION sources are highly compact.

\begin{figure}
    \centering
    \includegraphics[width=\textwidth]{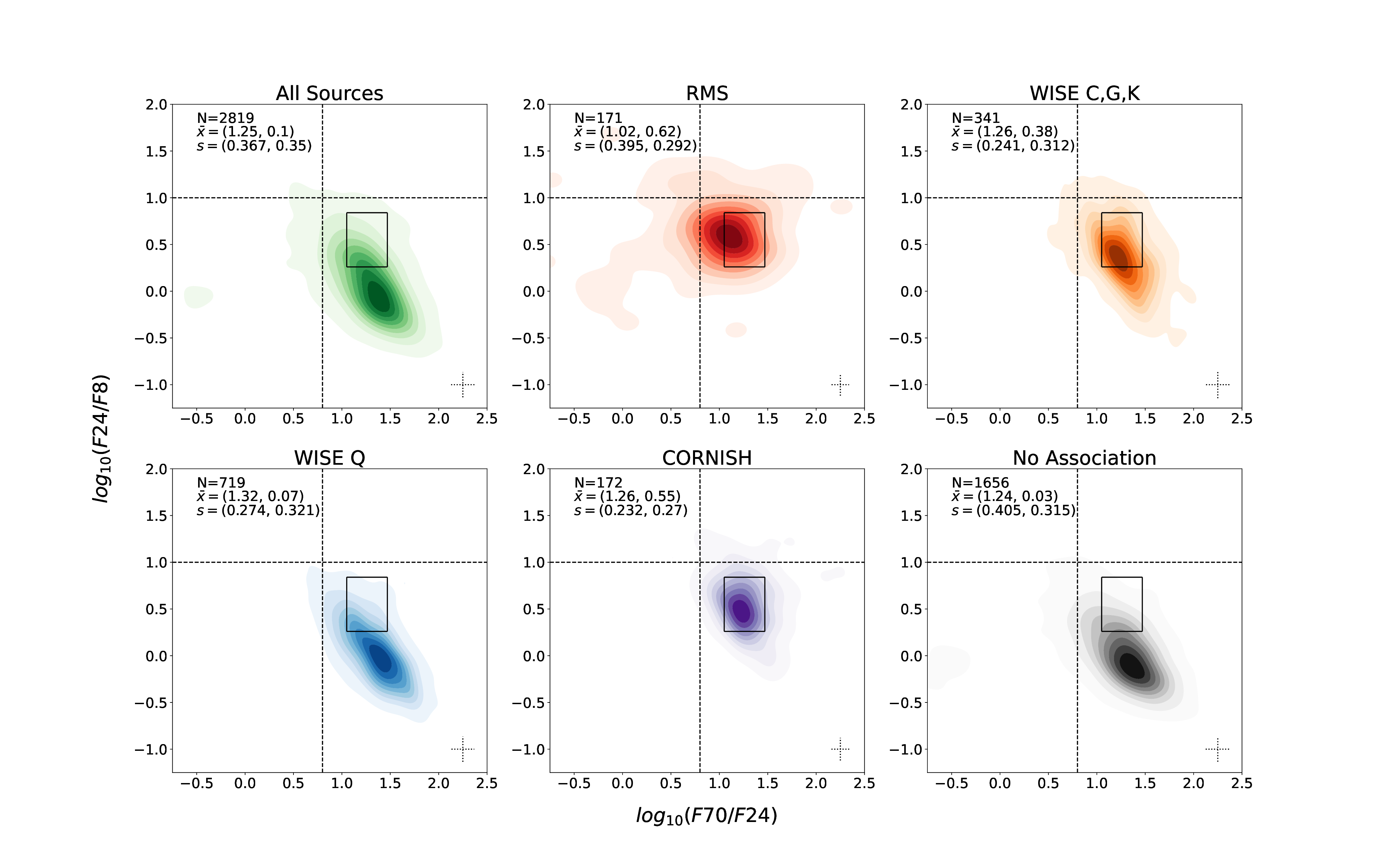}
    \caption{$\log_{10}(F_{24}/F_8)$ vs. $\log_{10}(F_{70}/F_{24})$ color-color kernel density estimate plots created using the photometry data for MIRION catalog sources. Darker colors indicate higher source density in that region. The plots shown exclude sources flagged for Poor Confidence or No Obvious Source, or with a fractional error $>0.5$ at 70~\um, 24~\um, or 8~\um. Separate color-color plots were created based on MIRION source crossmatches with other catalogs; the top left plot includes all sources, while the bottom right shows sources with no crossmatch in the RMS, WISE, or CORNISH catalogs. The number of sources plotted ($N$), average $\log_{10}(F_{70}/F_{24})$ and $\log_{10}(F_{24}/F_8)$ colors ($\bar{x}$), and standard deviations of the $\log_{10}(F_{70}/F_{24})$ and $\log_{10}(F_{24}/F_8)$ colors ($s$) are shown in the top left corner for each plot, and the dotted crosshatches in the bottom right corner show the average uncertainties. The dashed lines are the \HII-PNe cutoffs from \cite{Anderson2012}, and the solid rectangular region indicates the average \HII region colors from \cite{Anderson2012}.}
    \label{fig:ccp}
\end{figure}

IR $\log_{10}(F_{24}/F_8)$ vs. $\log_{10}(F_{70}/F_{24})$ color-color plots are shown in Figure \ref{fig:ccp}. These plots include only sources with no flags for Poor Confidence or No Obvious Source and a fractional error below the 0.5 cut-off, giving a total of 2819 sources. This is an order of magnitude higher than the number of sources analyzed similarly in the WKD21 pilot region (219 sources). In general, the findings using the entire MIRION catalog are consistent with the WKD21 pilot region results. MIRION sources with RMS, CORNISH, or WISE C/G/K counterparts are centered in the average \HII~color regions from \citet{Anderson2012}. This is unsurprising since most of these sources are indeed \HII~regions. In contrast, MIRION sources with a WISE Q association or no crossmatches in the WISE, RMS, or CORNISH catalogs (``No Association" in Figure \ref{fig:ccp}) are shifted toward lower $\log_{10}(F_{24}/F_8)$. As described in WKD21, higher $\log_{10}(F_{24}/F_8)$ values may serve as an indicator of PAH destruction by hard UV emission, while lower $\log_{10}(F_{24}/F_8)$ values reflect excitation, but not destruction, of PAHs by a softer UV field. MIRION sources with lower $\log_{10}(F_{24}/F_8)$ values are consistent with the expected color trends of intermediate-mass star-forming regions, although some of these sites may eventually produce high-mass stars. Since $60\%$ of the sources included in the Figure \ref{fig:ccp} analysis are ``No Association" sources, and an additional $25\%$ are matched with WISE Q sources, these color-color plots further demonstrate that a majority of MIRION sources are associated with candidate intermediate-mass SFRs.

MIRION sources with lower $\log_{10}(F_{70}/F_{24})$ values in Figure \ref{fig:ccp} may reflect cooler stellar environments with SEDs that peak at longer wavelengths. The mean value of this ratio does not change significantly across the subgroups plotted in the figure, indicating the dust temperature in these objects' environments is similar. Longer wavelength observations spanning the greybody curve of these regions as done by \citet{Elia2017} would better constrain the temperature.

The 0.5 fractional error cut-off used in this analysis was chose based on preference and to be consistent with the analysis in WKD21. MIRION catalog users can sort and analyze the data based on their own fractional error cut-off choices. To explore the impact of the fractional error cut-off on our conclusions, we examined the distributions shown in Figures~\ref{fig:colorhist} and \ref{fig:ccp} with fractional error cut-offs of 0.3 and 0.7. When the cut-off is adjusted to either of these values, the mean colors shown in Figure \ref{fig:colorhist}  change by $<0.02$, while the mean colors shown in Figure \ref{fig:ccp}, change by $<0.05$. Varying the fractional error cut-off within this range does not fundamentally change the results of our analysis.

\section{Summary and Conclusions} \label{sec:conclusions} 

In this work, we presented the MIRION catalog, which contains infrared fluxes, velocities, distances, and catalog crossmatches for all $\sim6000$ sources identified as YBs by MWP volunteers, as well as physical properties of sources associated with entries in the Hi-GAL CSC. MIRION catalog entries are heterogeneous and represent a range of objects that share the ``yellow" MIR-color signature, which led the MWP volunteers to mark them for inclusion in the catalog. Thus, we have replaced the term ``yellowballs (YBs)'' with ``MIRION sources,'' to more accurately reflect the diverse physical nature of these objects.

We have demonstrated that the vast majority of all MIRION sources are compact SFRs. Most are consistent with accreting protoclusters that span a wide range in mass ($0 < \log_{10}(M/\mathrm{M_\sun}) < 4$) and luminosity ($1 < \log_{10}(L/\mathrm{L_\sun}) < 5$). Over 50\% of MIRION sources may highlight sites of intermediate-mass star formation; however, given uncertainties related to the details of clump evolution we cannot rule out the possibility that some of these sites may eventually produce high-mass stars. This study has included more than an order of magnitude more sources than the WKD21 study in its analysis. Of particular significance is the identification of approximately an order of magnitude more sources associated with 10-100 $\mathrm{M_\sun}$ clumps than were found in the WKD21 study. 
%[GWC - Quantify? How many in this range?
%CK - I don't think we need to quantify here - I do mention
% numbers now back in 6.1] 
%KD - 6.1 doesn't give a numerical comparison to WKD21- eyeballing WKD21 Fig 4 looks like 10s of sources in tht region so I went with "about an order of magnitude"

Additionally, our large database of infrared flux measurements has enabled us to greatly expand our color-based analysis. The vast majority of MIRION sources have lower $\log_{10}(F_{12}/F_{8})$ colors than typical \HII regions. Together with the finding that MIRION sources with and without signatures of high-mass star formation (\HII regions and high-mass YSOs) have no significant size difference, this suggests that most MIRION sources delineate compact PDRs with high levels of PAH ionization.

Color-color plots indicate a significant $\log_{10}(F_{24}/F_8)$ color difference between sources with and without signatures of high-mass star formation. While MIRION sources with RMS, CORNISH, and WISE C/G/K counterparts occupy the color space associated with \HII regions, those with WISE Q and no associations (comprising the majority of all sources) have significantly lower $\log_{10}(F_{24}/F_8)$ colors. This result suggests that these sources identify softer UV environments in which PAH emitters are excited but not destroyed, consistent with expected characteristics of intermediate-mass star formation. Both samples have similar $\log_{10}(F_{70}/F_{24})$ colors, which is consistent with a very slight difference in the median dust temperature between these samples. We conclude that the sources currently lacking indicators of high-mass star formation represent a mix of intermediate-mass star-forming environments and very young regions that may eventually form high-mass stars. We will utilize the MIRION catalog to probe the relationship between IR colors and the evolutionary stage and physical properties of star-forming clumps in a subsequent paper.

\begin{acknowledgments}

The authors wish to thank Sean Carey and Alberto Noriega-Crespo for useful discussions regarding MIPSGAL data. We thank Matt Povich for his work as PI of the MWP, which generated the original yellowball database. We are especially grateful to undergraduate astronomy students who beta-tested the PERYSCOPE materials and contributed their photometric results; these students are listed in Table \ref{tab:peryscope}.

 Work conducted at The College of Idaho was supported by the Murdock Charitable Trust (Grants NS-2016246, SR-201811723, and SR-202119904). Authors KD, GWC, CK, NL, MC, and EB were supported by NSF Grant No. 2307806. The PERYSCOPE Project development and beta testing was supported by a NASA Citizen Science Seed Funding Grant (Grant No. 20-CSSFP20-001).  

This work is based in part on observations made with the Spitzer Space Telescope, which was operated by the Jet Propulsion Laboratory, California Institute of Technology under a contract with NASA. This publication makes use of data products from the Wide-field Infrared Survey Explorer, which is a joint project of the University of California, Los Angeles, and the Jet Propulsion Laboratory/California Institute of Technology, funded by the National Aeronautics and Space Administration. Herschel is an ESA space observatory with science instruments provided by European-led Principal Investigator consortia and with important participation from NASA. This research has made use of the VizieR catalogue access tool, CDS, Strasbourg Astronomical Observatory, France \citep{vizier}. This research made use of ds9, a tool for data visualization supported by the Chandra X-ray Science Center (CXC) and the High Energy Astrophysics Science Archive Center (HEASARC) with support from the JWST Mission office at the Space Telescope Science Institute for 3D visualization.

%This research has made use of the VizieR catalogue access tool, CDS, Strasbourg, France. This research made use of pandas \citep{McKinney_2010, McKinney_2011} This research made use of Astropy, a community-developed core Python package for Astronomy \citep{2018AJ....156..123A, 2013A&A...558A..33A} This research made use of SciPy \citep{Virtanen_2020} This research made use of matplotlib, a Python library for publication quality graphics \citep{Hunter:2007} This research made use of ds9, a tool for data visualization supported by the Chandra X-ray Science Center (CXC) and the High Energy Astrophysics Science Archive Center (HEASARC) with support from the JWST Mission office at the Space Telescope Science Institute for 3D visualization. This research made use of NumPy \citep{harris2020array} This research made use of TOPCAT, an interactive graphical viewer and editor for tabular data \citep{2005ASPC..347...29T}  This publication makes use of data products from the Wide-field Infrared Survey Explorer\citep{2010AJ....140.1868W}, which is a joint project of the University of California, Los Angeles, and the Jet Propulsion Laboratory/California Institute of Technology, funded by the National Aeronautics and Space Administration. 

\begin{deluxetable*}{cccc}
\tablewidth{0pt}
\tablecaption{PERYSCOPE Participants and Beta-Testers \label{tab:peryscope}}
\tablehead{
\colhead{ } & \colhead{ } & \colhead{ } & \colhead{ }
}
\startdata
 Santosh Acharya                  & Tyler Carlson & Bradley Mathew Cushman & Allison Van Dyke   \\
 Kat Falk                         & Brielle Fieuw & Noah Ghiselli          & Caden Handran      \\
 Jaden Lynn Hernandez             & Lex Jones     & Brady Joyner           &  
 Madeline Khoury  \\         Whitt Miller  & Samaje Morgan          &        
 Royce Samaniego        &          Lily Schlake  \\ Asta Shakti Suman Sharma  & E.T. Sherwin  &         Ian Madsen Stowman &
 Rebecca Amaya Villarreal-Mendoza \\ Anna Willcuts & JD Willis  &          &  \\
\enddata
\end{deluxetable*}

\end{acknowledgments}


\vspace{5mm}
%\facilities{}


\software{TopCat \citep{topcat}, Astropy \citep{astropy1, astropy2}, Matplotlib \citep{matplotlib}, NumPy \citep{numpy}, pandas \citep{McKinney_2010, McKinney_2011}, SciPy \citep{Virtanen_2020}.}

\bibliography{YBCat}{}
\bibliographystyle{aasjournal}



\end{document}